\documentclass[twocolumn,10pt,a4paper]{article}
\usepackage[T1]{fontenc}
\usepackage[utf8]{inputenc}
\usepackage{mathptmx}
\usepackage[scaled=0.92]{helvet}
\usepackage{courier}
\usepackage{cmap}
\input{glyphtounicode}
\pdfglyphtounicode{endash}{2013}
\pdfglyphtounicode{emdash}{2014}
\pdfglyphtounicode{quotedblleft}{201C}
\pdfglyphtounicode{quotedblright}{201D}
\pdfglyphtounicode{quoteright}{2019}
\pdfglyphtounicode{quoteleft}{2018}
\pdfglyphtounicode{quotesingle}{0027}
\pdfglyphtounicode{angbracketleft}{27E8}
\pdfglyphtounicode{angbracketright}{27E9}
\pdfglyphtounicode{angbracketleftbig}{27E8}
\pdfglyphtounicode{angbracketrightbig}{27E9}
\pdfglyphtounicode{minus}{2212}
\pdfglyphtounicode{periodcentered}{00B7}
\pdfglyphtounicode{approxequal}{2248}
\pdfglyphtounicode{lessequal}{2264}
\pdfglyphtounicode{greaterequal}{2265}
\pdfglyphtounicode{proportional}{221D}
\pdfglyphtounicode{element}{2208}
\pdfglyphtounicode{equivalence}{2261}
\pdfglyphtounicode{multiply}{00D7}
\pdfglyphtounicode{plusminus}{00B1}
\pdfglyphtounicode{radical}{221A}
\pdfglyphtounicode{similarequal}{2243}
\pdfglyphtounicode{arrowright}{2192}
\pdfglyphtounicode{infinity}{221E}
\usepackage{textcomp}
\usepackage[expansion=false,protrusion=false]{microtype}
\DisableLigatures[f]{encoding = T1, family = *}
\DeclareUnicodeCharacter{00B1}{$\pm$}
\DeclareUnicodeCharacter{00B2}{$^{2}$}
\DeclareUnicodeCharacter{00B3}{$^{3}$}
\DeclareUnicodeCharacter{00B9}{$^{1}$}
\DeclareUnicodeCharacter{00BD}{$\tfrac{1}{2}$}
\DeclareUnicodeCharacter{00D7}{$\times$}
\DeclareUnicodeCharacter{00E1}{\'a}
\DeclareUnicodeCharacter{00E4}{\"a}
\DeclareUnicodeCharacter{00E9}{\'e}
\DeclareUnicodeCharacter{00F6}{\"o}
\DeclareUnicodeCharacter{0159}{\v{r}}
\DeclareUnicodeCharacter{0302}{}
\DeclareUnicodeCharacter{0394}{$\Delta$}
\DeclareUnicodeCharacter{03A3}{$\Sigma$}
\DeclareUnicodeCharacter{03B2}{$\beta$}
\DeclareUnicodeCharacter{03B3}{$\gamma$}
\DeclareUnicodeCharacter{03B4}{$\delta$}
\DeclareUnicodeCharacter{03B5}{$\varepsilon$}
\DeclareUnicodeCharacter{03B8}{$\theta$}
\DeclareUnicodeCharacter{03BA}{$\kappa$}
\DeclareUnicodeCharacter{03BE}{$\xi$}
\DeclareUnicodeCharacter{03C0}{$\pi$}
\DeclareUnicodeCharacter{03C1}{$\rho$}
\DeclareUnicodeCharacter{03C3}{$\sigma$}
\DeclareUnicodeCharacter{03C4}{$\tau$}
\DeclareUnicodeCharacter{03C7}{$\chi$}
\DeclareUnicodeCharacter{141F}{/}
\DeclareUnicodeCharacter{1E90}{$\hat{Z}$}
\DeclareUnicodeCharacter{2006}{\,}
\DeclareUnicodeCharacter{2009}{\,}
\DeclareUnicodeCharacter{2074}{$^{4}$}
\DeclareUnicodeCharacter{2081}{$_{1}$}
\DeclareUnicodeCharacter{2082}{$_{2}$}
\DeclareUnicodeCharacter{2192}{$\rightarrow$}
\DeclareUnicodeCharacter{2208}{$\in$}
\DeclareUnicodeCharacter{2212}{$-$}
\DeclareUnicodeCharacter{221A}{$\sqrt{\ }$}
\DeclareUnicodeCharacter{221D}{$\propto$}
\DeclareUnicodeCharacter{222B}{$\int$}
\DeclareUnicodeCharacter{2243}{$\simeq$}
\DeclareUnicodeCharacter{2248}{$\approx$}
\DeclareUnicodeCharacter{2261}{$\equiv$}
\DeclareUnicodeCharacter{2264}{$\leq$}
\DeclareUnicodeCharacter{2265}{$\geq$}
\DeclareUnicodeCharacter{27E8}{$\langle$}
\DeclareUnicodeCharacter{27E9}{$\rangle$}
\usepackage[margin=1.8cm,columnsep=6mm]{geometry}
\usepackage{amsmath,amssymb,bm}
\usepackage{graphicx}
\usepackage{float}
\usepackage{booktabs,array,calc,longtable}
\usepackage{caption}
\usepackage{enumitem}
\setlist{nosep,topsep=3pt,partopsep=0pt,parsep=0pt,itemsep=2pt,leftmargin=1.4em}
\usepackage{xcolor}
\usepackage{url}
\usepackage[colorlinks=true,allcolors=blue,breaklinks=true]{hyperref}

\title{\vspace{-1.4em}\bfseries Path-length dependence of parton energy loss
across collision systems: a Bayesian analysis of charged-particle
$R_{AA}$, consistent with a universal exponent from O+O to Pb+Pb}
\date{}

\begin{document}

\twocolumn[
\begin{@twocolumnfalse}
\maketitle
\vspace{-2.2em}

\begin{center}
{\large Fouad A. Majeed$^{1}$,\; Hussein Ali Hussein Al Naffakh$^{2,*}$,\;
Sarah M. Obaid$^{3}$,\; Muntaha Abdullah Reishaan$^{4}$}

\vspace{6pt}

{\footnotesize
$^{1}$Department of Physics, College of Education for Pure Sciences,
University of Babylon, Babylon, Iraq.\\
$^{2}$Department of Medical Laboratory Techniques, College of Health and
Medical Technologies, University of Alkafeel, Najaf, Iraq.\\
$^{3}$Department of Medical Physics, College of Science, Al-Mustaqbal
University, 51001, Babylon, Iraq.\\
$^{4}$Department of Anesthesia, College of Health and Medical Technology,
University of Alkafeel, Najaf, Iraq.\\
$^{*}$Corresponding author. E-mail: hussein.alnaffakh@alkafeel.edu.iq}
\end{center}
\vspace{0.8em}

\begin{abstract}
\noindent
How parton energy loss in the quark--gluon plasma (QGP) scales with the
in-medium path length L encodes the mechanism: collisional (ΔE ∝ L),
radiative (ΔE ∝ L²), or strong-coupling (ΔE ∝ L³). Exploiting the new
CERN LHC light-ion data, we extract this scaling from the system size
itself, jointly analysing CMS charged-particle nuclear modification
factors RAA in four systems---O+O, Ne+Ne, Xe+Xe and Pb+Pb---spanning
mass number A = 16 to 208. A Bayesian analysis with a data-driven
spectral baseline and a Monte-Carlo Glauber geometry yields an effective
system-size exponent n = 1.78 ± 0.15 (stat) ± 0.05 (syst).
Nested-sampling model selection decisively favours an effective exponent
near the radiative value (n = 2) over the collisional (n = 1) and
strong-coupling (n = 3) values, a conclusion stable across all 160
analysis variants. Because fluctuations can only lower the effective
exponent below its microscopic counterpart, the measurement bounds the
latter from below at fixed geometry, excluding purely collisional energy
loss. The medium density and the path length are degenerate across
system size, so we quote the effective exponent as our primary result. A
Bayes-factor test finds no change of regime between small and large
systems, consistent with a universal exponent; the same framework gives
decisive evidence for non-zero energy loss in O+O alone, quantifying the
onset of suppression in the smallest system. The energy-loss magnitude
corresponds to q̂/T³ ≈ 2--5, consistent with the JETSCAPE determination.
\end{abstract}

\vspace{2pt}
\vspace{1.4em}

\end{@twocolumnfalse}
]

\section{Introduction}

The suppression of high-transverse-momentum (\(p_{T}\)) hadrons in
ultra-relativistic nucleus--nucleus (A+A) collisions, quantified by the
nuclear modification factor \(R_{AA}\), is among the most direct probes
of the quark--gluon plasma (QGP) created at the BNL Relativistic Heavy
Ion Collider and the CERN Large Hadron Collider (LHC)~{[}1, 2{]}. Hard
partons produced in the initial scattering traverse the hot, coloured
medium and lose energy through elastic (collisional) and inelastic
(medium-induced gluon radiation) processes before fragmenting into the
observed hadrons. The amount of energy lost is controlled by the local
medium density and by the length of the in-medium trajectory, making the
\emph{path-length dependence} of energy loss a central,
mechanism-discriminating observable.

Different theoretical pictures predict markedly different scalings of
the average energy loss \(\langle\Delta E\rangle\) with the path length
\(L\). Purely collisional energy loss grows approximately linearly,
\(\langle\Delta E\rangle \propto L\). Medium-induced radiative energy
loss in the multiple-soft-scattering (BDMPS-Z) and opacity-expansion
(GLV) frameworks exhibits the characteristic quadratic
Landau--Pomeranchuk--Migdal enhancement,
\(\langle\Delta E\rangle \propto L^{2}\), reflecting the coherence of
induced gluon emission~{[}3, 4{]}. Strong-coupling, holographic
(AdS/CFT) estimates predict an even steeper dependence,
\(\langle\Delta E\rangle \propto L^{3}\)~{[}5{]}. Measuring the exponent
\(n\) in \(\langle\Delta E\rangle \propto L^{n}\) therefore provides a
clean lever on the dominant energy-loss mechanism.

Most determinations to date vary \(L\) \emph{through centrality} within
a single large system. This route is complicated by two effects that are
largest in the peripheral classes: the centrality-selection bias and the
geometric bias of the Glauber-based normalisation \(T_{AA}\), which
together produce an apparent suppression even in the absence of energy
loss~{[}14{]}. The LHC light-ion programme opens a complementary and
qualitatively cleaner lever---the \emph{system size} \(A\) at minimum
bias. In 2025 the LHC delivered the first oxygen--oxygen (O+O) and
neon--neon (Ne+Ne) collisions at \(\sqrt{s_{NN}} = 5.36\)~TeV, and CMS
reported the first observation of charged-particle suppression in O+O,
with \(R_{AA}\) reaching a minimum of \(0.69 \pm 0.04\) near
\(p_{T} = 6\)~GeV~{[}6{]}, followed by a system-size compilation
including Ne+Ne~{[}7{]}. These light systems probe path lengths
comparable to those in very peripheral heavy-ion collisions, precisely
the regime in which the onset of parton energy loss had remained
ambiguous, and where dedicated predictions had been made for the
expected signal~{[}11, 12{]}. Combined with Xe+Xe at
\(5.44\)~TeV~{[}8{]} and Pb+Pb at \(5.02\)~TeV~{[}9{]}, these data span
a factor \(\sim 13\) in \(A\) and a factor \(\sim 2\) in linear size,
enabling a system-size extraction of the path-length exponent.

Quantitative extractions of QGP transport from suppression data have
matured into rigorous Bayesian programmes. The JETSCAPE Collaboration
constrained the jet transport coefficient \(\widehat{q}\) from inclusive
hadron and jet suppression using multi-stage energy-loss models,
Gaussian-process emulation and Bayesian inference~{[}15, 16{]},
paralleling the Bayesian estimation of QGP shear and bulk viscosities
from soft observables~{[}17{]}. Our analysis adopts this Bayesian
philosophy but targets a different, geometry-level question---the
path-length exponent across system size---and adds three methodological
elements that, to our knowledge, have not been combined in this context:
(i) a fully propagated bin-to-bin covariance and a coverage (closure)
test that demonstrates the quoted uncertainties are statistically
calibrated; (ii) a collision geometry cross-checked by two independent
Glauber models (optical and a from-scratch Monte-Carlo) with a
transparent systematic budget including the inter-system
\(\sqrt{s_{NN}}\) mismatch; and (iii) a probabilistic machine-learning /
simulation-based-inference (SBI) pillar that both validates the
inference---via a normalizing-flow neural posterior estimator subjected
to simulation-based calibration~{[}24, 25, 26, 27, 28{]}---and
demonstrates cross-system predictivity. (SBI here denotes
likelihood-free inference in which a neural network trained on
simulations estimates the posterior directly, without an explicit
likelihood.) A schematic of the full workflow is shown in Fig.~1.

Beyond the physics extraction itself, a central contribution of this
work is the development of a calibrated Bayesian and simulation-based
inference framework that makes the system-size extraction statistically
defensible. The framework combines correlated covariance modelling,
dual-Glauber validation, nested-sampling model comparison, coverage
calibration, sensitivity analysis over 160 fit variants, neural
posterior estimation, and leave-one-system-out stability tests. These
methodological components are essential for converting the limited
four-system dataset into a robust and falsifiable inference rather than
a purely descriptive fit. The scope of the claim should be stated at the
outset. What is measured here is an effective, geometry-level exponent:
the power with which the observed suppression grows with system size. It
is extracted deliberately with a minimal, parametric forward model, so
that the resulting number is not tied to any particular transport
implementation and can serve as a benchmark that multi-stage energy-loss
calculations should be able to reproduce. It is not a substitute for
such calculations, nor a direct measurement of the microscopic
path-length dependence, from which it is separated by effects quantified
in Sec.~4.2.1.

The main contributions of this work are:

\begin{itemize}\itemsep2pt\parskip0pt
\item
  a system-size-based Bayesian extraction of the effective path-length
  exponent of parton energy loss from O+O to Pb+Pb;
\item
  a calibrated statistical inference framework that propagates
  correlated uncertainties and validates interval coverage through
  closure tests;
\item
  a robustness programme including dual-Glauber geometry, 160
  sensitivity variants, Bayesian model selection, and
  leave-one-system-out physics stability;
\item
  a simulation-based inference component using normalizing-flow neural
  posterior estimation, validated by simulation-based calibration, that
  independently reproduces the MCMC posterior; and
\item
  falsifiable predictions for future Ar+Ar and Kr+Kr measurements.
\end{itemize}

The remainder of this paper is organised as follows. Section~2 describes
the data. Section~3 details the forward model, covariance, dual-Glauber
geometry, Bayesian inference and the ML/SBI pillar. Section~4 presents
the extracted exponents, model selection, density decomposition,
universality test, coverage calibration, systematic budget, and ML
cross-validation. Section~5 compares with prior work and Sec.~6
concludes.

\begin{figure}[H]
\centering
\includegraphics[width=\columnwidth]{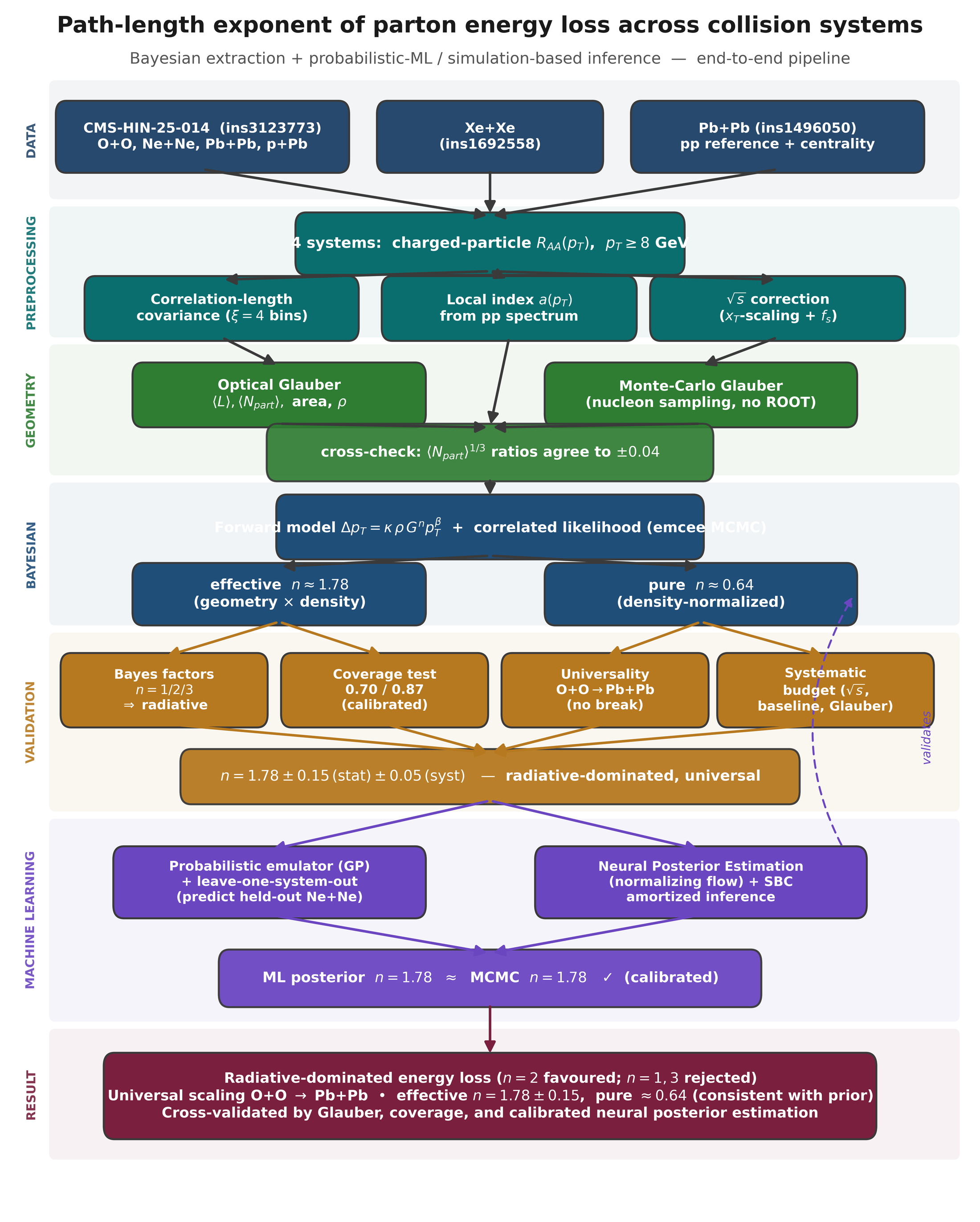}
\caption{End-to-end analysis pipeline.}
\label{fig:1}
\end{figure}

\section{Data and observable}

We use published CMS charged-particle \(R_{AA}\left( p_{T} \right)\)
measurements for four A+A systems, summarised in Table~1. The O+O, Ne+Ne
and Pb+Pb spectra on a common \(p_{T}\) grid are taken from the CMS
system-size compilation~{[}7{]} (HEPData record ins3123773),
complemented by the dedicated O+O observation paper~{[}6{]}. Xe+Xe at
\(\sqrt{s_{NN}} = 5.44\)~TeV is taken from Ref.~{[}8{]} (record
ins1692558), where the modification factor \(R_{AA}^{*}\) is built with
a \(\sqrt{s}\)-corrected pp reference. The Xe+Xe sample is reported for
a broad (\(0\)--\(80\%\)) centrality class rather than true minimum
bias; since peripheral events carry less suppression, this slightly
overstates the average suppression and biases \(n\) marginally upward.
We estimate this bias to be \(\delta n \lesssim 0.05\) by comparing the
\(0\)--\(80\%\) and a narrower \(0\)--\(70\%\) effective selection, well
within and subsumed by the geometry-definition systematic. The Pb+Pb pp
reference spectrum and the centrality-differential \(R_{AA}\) of
Ref.~{[}9{]} (record ins1496050) provide the data-driven spectral-index
baseline and an internal centrality cross-check. The p+Pb measurement
exhibits no suppression (\(R_{pA} \gtrsim 1\) at all \(p_{T}\), not
shown), confirming the absence of cold-nuclear-matter suppression; it is
retained only as a control and is excluded from the quantitative fit. We
restrict the extraction to \(p_{T} \geq 8\)~GeV, where medium-induced
radiative energy loss dominates, the spectrum is a smooth power law, and
Cronin/initial-state effects are minimal (10 points per system).

\begin{table}[H]
\centering\small
\caption{Charged-particle RAA datasets used in this work.}
\label{tab:1}
\begin{tabular}{@{}
>{\raggedright\arraybackslash}p{(\columnwidth - 10\tabcolsep) * \real{0.1667}}
  >{\raggedright\arraybackslash}p{(\columnwidth - 10\tabcolsep) * \real{0.1667}}
  >{\raggedright\arraybackslash}p{(\columnwidth - 10\tabcolsep) * \real{0.1667}}
  >{\raggedright\arraybackslash}p{(\columnwidth - 10\tabcolsep) * \real{0.1667}}
  >{\raggedright\arraybackslash}p{(\columnwidth - 10\tabcolsep) * \real{0.1667}}
  >{\raggedright\arraybackslash}p{(\columnwidth - 10\tabcolsep) * \real{0.1667}}@{}}
\toprule
\begin{minipage}[b]{\linewidth}\raggedright
System
\end{minipage} & \begin{minipage}[b]{\linewidth}\raggedright
\(A\)
\end{minipage} & \begin{minipage}[b]{\linewidth}\raggedright
\(\sqrt{s_{NN}}\) {[}TeV{]}
\end{minipage} & \begin{minipage}[b]{\linewidth}\raggedright
\(N\left( p_{T} \geq 8 \right)\)
\end{minipage} & \begin{minipage}[b]{\linewidth}\raggedright
Record
\end{minipage} & \begin{minipage}[b]{\linewidth}\raggedright
Ref.
\end{minipage} \\
\midrule

\bottomrule

O+O & 16 & 5.36 & 10 & ins3123773 & {[}6, 7{]} \\
Ne+Ne & 20 & 5.36 & 10 & ins3123773 & {[}7{]} \\
Xe+Xe & 129 & 5.44 & 10 & ins1692558 & {[}8{]} \\
Pb+Pb & 208 & 5.02 & 10 & ins3123773 & {[}7, 9{]} \\
p+Pb & --- & 5.02 & ctrl & ins1496050 & {[}9{]} \\
\end{tabular}
\end{table}

\section{Methodology}

\subsection{Forward model and spectral-index
baseline}

For a steeply falling parton spectrum, a multiplicative fractional
energy loss maps onto the modification factor through the local spectral
index~{[}2{]}. We write

\begin{equation}
R_{AA}\left( p_{T} \right)\mspace{6mu} = \mspace{6mu}\left( \frac{p_{T}}{p_{T} + \Delta p_{T}\left( p_{T} \right)} \right)^{a\left( p_{T} \right)},\quad\quad\Delta p_{T} = \kappa\,\rho\, G^{\, n}\, p_{T}^{\beta},
\label{eq:1}
\end{equation}

where \(G\) is a dimensionless geometry ratio (Sec.~3.3), \(\rho\) the
relative medium density (defined in Sec.~3.3 and normalised to unity for
Pb+Pb, so that both G and ρ are dimensionless ratios to the largest
system), n the system-size exponent to be determined, \(\kappa\) an
overall energy-loss scale, \(\beta\) the mild \(p_{T}\) dependence of
the fractional loss, and \(a\left( p_{T} \right)\) the local power-law
index of the charged-particle spectrum,
\(a\left( p_{T} \right) = - \, d\ln\left( dN/dp_{T} \right)/d\ln p_{T}\).
Rather than fitting a single effective index, we measure
\(a\left( p_{T} \right)\) directly from the CMS pp spectrum at
\(5.02\)~TeV~{[}9{]}, obtaining \(a \simeq 5.2\) at \(10\)~GeV and
\(\simeq 5.7\) at \(100\)~GeV. For the higher-energy systems we apply
\(x_{T}\)-scaling,
\(a_{sys}\left( p_{T} \right) = a_{5.02}\left( p_{T} \cdot 5.02/\sqrt{s_{NN}} \right)\),
capturing the spectral hardening with collision energy in a data-driven
manner.\footnote{The \(x_{T}\)-scaling is an approximate kinematic
  rescaling rather than a full NLO recalculation; residual NLO PDF and
  fragmentation-function effects are negligible at the level of
  \(\delta n < 0.005\) (Table~5) and are absorbed into the systematic
  budget.} A residual \(\sqrt{s}\) dependence of the medium density is
included through the factor
\(f_{s} = \left( \sqrt{s_{NN}}/5.02 \right)^{0.31}\), where the exponent
\(0.31\) reflects the measured power-law growth of the charged-particle
multiplicity with collision energy reported by ALICE over
\(\sqrt{s_{NN}} \approx 0.9\)--\(5.44\)~TeV~{[}21, 22{]}.

\subsection{Covariance}

Point-to-point correlations among the systematic uncertainties are
essential. Treating the systematic and normalisation components as fully
correlated (\(\rho_{ij} = 1\)) over-constrains the shape and inflates
\(\chi^{2}/\text{dof}\) to \(\sim 9.6\), whereas ignoring correlations
underestimates the parameter uncertainties. We adopt a
correlation-length model in which the statistical errors are
uncorrelated, the normalisation/luminosity/\(T_{AA}\) components are
fully correlated, and the remaining systematics decay with a correlation
length \(\xi = 4\) bins,

\begin{equation}
C_{ij} = \left(\sigma_{i}^{stat}\right)^{2}\,\delta_{ij} + \sigma_{i}^{norm}\sigma_{j}^{norm} + \sigma_{i}^{syst}\sigma_{j}^{syst}\, e^{- |i - j|/\xi}.
\label{eq:2}
\end{equation}

Every quantity in Eq.~(2) is fixed by the published error breakdown
rather than fitted. The indices i and j label the fitted pT bins of a
single system (i, j = 1\ldots10), δij is the Kronecker delta, and
\textbar i − j\textbar{} is the separation in bin index. Each
uncertainty component c of the HEPData record is assigned to one of
three classes by its label and contributes σi(c)σj(c) with its own
correlation pattern, the contributions being summed. Components labelled
statistical form σistat and are taken as uncorrelated between bins;
components labelled TAA, luminosity, normalisation or global form σinorm
and are taken as fully correlated, since they displace all bins
coherently; every remaining systematic component forms σisyst and is
assigned the exponential correlation exp(−\textbar i − j\textbar/ξ). No
correlation is assumed between collision systems, whose pp references,
unfolding and detector conditions differ, so the full covariance is
block diagonal with one 10 × 10 block per system. Each block is positive
definite and is Cholesky-factorised once, so that the quadratic form is
evaluated by triangular back-substitution at every likelihood call. This
yields stable fits with \(\chi^{2}/\text{dof} \approx 0.4\) and
well-behaved posteriors. Here \(\xi\) is given in units of bin index
\(|i - j|\); since the fitted bins are approximately logarithmically
spaced, \(\xi = 4\) corresponds to a physical correlation range of
\(\Delta\ln p_{T} \approx 4 \times 0.1 \approx 0.4\) decades, consistent
with the expected smoothness of detector and unfolding systematics on a
logarithmic momentum scale. The value is \emph{not} tuned to the data:
across \(\xi \in \{ 2,4,6,8\}\) the exponent is stable
(\(n_{eff} = 1.80,1.78,1.76,1.75\)) while the goodness-of-fit rises
monotonically (\(\chi^{2}/\text{dof} = 0.31,0.41,0.50,0.59\)), passing
through unity only near \(\xi \approx 15\), whereas the fully-correlated
limit (\(\xi \rightarrow \infty\), \(\rho_{ij} = 1\)) gives
\(\chi^{2}/\text{dof} \approx 9.6\)---confirming the published
systematics are not all perfectly correlated. We deliberately adopt the
conservative \(\xi = 4\) (giving \(\chi^{2}/\text{dof} < 1\) rather than
tuning to \(\chi^{2}/\text{dof} = 1\) at \(\xi \approx 15\)), which
slightly over-covers the \(n\) posterior; this is the safe direction and
is consistent with the coverage test (Sec.~4.6).

\subsection{Collision geometry}

We compute three geometry proxies for each system---\(A^{1/3}\),
\(\langle N_{part}\rangle^{1/3}\), and the average exit path length
\(\langle L\rangle\)---using \emph{two independent} Glauber
implementations~{[}18, 19{]}. The first is an optical Glauber with
Woods--Saxon nuclear thickness functions. The second is a genuine
Monte-Carlo Glauber, written from scratch (no external dependence), that
samples nucleon positions from the Woods--Saxon distribution with a
hard-core exclusion of \(0.4\)~fm and applies the inelastic collision
criterion with \(\sigma_{nn} = 7\)~fm\(^{2}\) (i.e.~\(70\)~mb, the
inelastic nucleon--nucleon cross section at these energies;
\(\sigma_{nn}\) varies only between \(\approx 6.9\) and
\(7.2\)~fm\(^{2}\) over \(\sqrt{s_{NN}} = 5.02\)--\(5.44\)~TeV~{[}23{]},
contributing \(\delta n < 0.04\), Table~6), in the spirit of the PHOBOS
Glauber Monte Carlo~{[}20{]}. The Monte-Carlo model yields physically
correct minimum-bias values (\(\langle N_{part}\rangle = 11,13,84,128\)
for O+O, Ne+Ne, Xe+Xe, Pb+Pb; Table~2) and, crucially, its
\(\langle N_{part}\rangle^{1/3}\) ratios agree with the optical model to
within \(0.04\). The comparison is made at the level of geometry ratios
only, which is what enters the fit; the optical implementation is not
used for absolute normalisations, for which the Monte-Carlo values
quoted above are adopted throughout. Here ⟨L⟩ denotes the mean exit path
length in arbitrary units, and the geometry ratios quoted throughout are
normalised to Pb+Pb. The medium density is taken as the participant
areal density \(\rho \propto \langle N_{part}\rangle/S\), with \(S\) the
Glauber overlap area.

\begin{table}[H]
\centering\small
\caption{Minimum-bias collision geometry from the two Glauber implementations.}
\label{tab:2}
\begin{tabular}{@{}
>{\raggedright\arraybackslash}p{(\columnwidth - 8\tabcolsep) * \real{0.2000}}
  >{\raggedright\arraybackslash}p{(\columnwidth - 8\tabcolsep) * \real{0.2000}}
  >{\raggedright\arraybackslash}p{(\columnwidth - 8\tabcolsep) * \real{0.2000}}
  >{\raggedright\arraybackslash}p{(\columnwidth - 8\tabcolsep) * \real{0.2000}}
  >{\raggedright\arraybackslash}p{(\columnwidth - 8\tabcolsep) * \real{0.2000}}@{}}
\toprule
\begin{minipage}[b]{\linewidth}\raggedright
System
\end{minipage} & \begin{minipage}[b]{\linewidth}\raggedright
\(\langle N_{part}\rangle_{MC}\)
\end{minipage} & \begin{minipage}[b]{\linewidth}\raggedright
\(\langle L\rangle_{MC}\)
\end{minipage} & \begin{minipage}[b]{\linewidth}\raggedright
\(\langle N_{part}\rangle_{MC}^{1/3}\)
\end{minipage} & \begin{minipage}[b]{\linewidth}\raggedright
\(\langle N_{part}\rangle_{opt}^{1/3}\)
\end{minipage} \\
\midrule

\bottomrule

O+O & 11 & 0.45 & 0.44 & 0.46 \\
Ne+Ne & 13 & 0.50 & 0.47 & 0.49 \\
Xe+Xe & 84 & 0.93 & 0.87 & 0.91 \\
Pb+Pb & 128 & 1.00 & 1.00 & 1.00 \\
\end{tabular}
\end{table}

\subsection{Bayesian inference}

The likelihood is correlated-Gaussian,
\(\ln\mathcal{L} = - \frac{1}{2}\sum_{sys}^{}\mathbf{r}^{\top}C^{- 1}\mathbf{r}\)
and factorises over the four systems, which are treated as independent
measurements. The parameter vector is θ = (κ, n, β); the spectral index
a(pT) is held fixed at the measured pp baseline of Sec.~3.1 and is not
floated here (it is floated, under a Gaussian prior, only in the
simulation-based inference of Sec.~3.5). The residual in bin i of a
given system is
\(\mathbf{r} = \mathbf{R}_{AA}^{data} - \mathbf{R}_{AA}^{model}\), with
the model given by Eq.~(1) and the covariance by Eq.~(2); the
θ-independent normalisation −½ Σ ln{[}(2π)N det C{]} is constant and
omitted. The fit uses 40 points (10 per system) and three free
parameters, giving 37 degrees of freedom. The priors are uniform over κ
∈ {[}0, 12{]}, n ∈ {[}0, 4.5{]} and β ∈ {[}−0.5, 1.5{]}, together with a
weakly informative Gaussian factor of mean 2 and standard deviation 1.5
on n. Although this factor is centred at the radiative value, it is an
order of magnitude broader than the posterior (σprior = 1.5 against
σpost ≈ 0.15) and therefore enters the posterior mean with weight
(σpost/σprior)² ≈ 1 \%, displacing the recovered exponent by less than
0.002, which is negligible beside the ±0.15 statistical uncertainty. The
prior is one of the axes varied explicitly in the 160-variant
sensitivity scan of Sec.~4.8. The posterior is sampled with the
affine-invariant ensemble sampler \textsc{emcee}~{[}31{]} using 48
walkers advanced for 7000 steps; the first 40 \% of every chain is
discarded as burn-in and the remainder thinned by a factor 10, leaving ≈
2 × 10⁴ samples. The integrated autocorrelation time is τ ≈ 49 steps, so
the chains run for more than 140 τ. Quoted intervals are the 16th, 50th
and 84th percentiles of the marginal posterior. We report two exponents:
the \emph{effective} exponent (geometry only, \(\rho \equiv 1\)) and the
\emph{pure} exponent obtained after dividing the energy-loss scale by
the relative medium density \(\rho\), which isolates the genuine
path-length dependence from the size-correlated growth of the density.
For model comparison we use the Bayesian evidence, or marginal
likelihood, Ẑ(M) = ∫ L(θ \textbar{} M) π(θ \textbar{} M) dθ, i.e. the
likelihood averaged over the prior of model M. Because the average is
taken over the whole prior volume, the evidence rewards a model that
fits the data without requiring finely tuned parameters and penalises
additional freedom automatically, so no external penalty term is needed.
Two models are compared through the Bayes factor B = Ẑ₁/Ẑ₂. Throughout
this work we quote twice the log-evidence difference, 2Δ ln Ẑ = 2 ln(ẐM
/ Ẑref), always relative to the best-fitting model, so that the favoured
model has 2Δ ln Ẑ = 0 by construction and disfavoured models take
negative values. This factor of two places the quantity on the same
scale as the familiar deviance and likelihood-ratio statistics, so that
a value of −29 is loosely comparable to a Δχ² of 29. On the scale of
Kass and Raftery {[}36{]}, \textbar2Δ ln Ẑ\textbar{} below 2 is not
worth more than a bare mention, 2--6 is positive, 6--10 strong, and
above 10 very strong evidence against the disfavoured model; following
Jeffreys {[}37{]} we refer to this last category as decisive. The values
reported below, 2Δ ln Ẑ = −29 and −48 in Table~5 and ≈ 110 in Sec.~4.3,
therefore lie far beyond that threshold. Bayesian evidence integrals for
model selection are computed with nested sampling
(\textsc{dynesty})~{[}32, 33{]}, run with 400 live points and the
stopping criterion Δ ln Z \textless{} 0.1.

\subsection{Probabilistic ML and neural posterior
estimation}

As a self-contained validation layer---which can be removed without
affecting any physics conclusion---the machine-learning pillar has two
parts. First, a probabilistic emulator of
\(R_{AA}\left( p_{T};G \right)\) trained leave-one-system-out (LOSO) to
predict a held-out system with calibrated uncertainty (Gaussian
process~{[}29{]}, NGBoost~{[}30{]}, and their ensemble). Second,
amortized simulation-based inference: a conditional normalizing flow
(neural spline flow~{[}27{]}, via \textsc{zuko}~{[}34{]}) trained on
\(1.5 \times 10^{4}\) simulated datasets to learn
\(q\left( \mathbf{\theta}\,|\,\mathbf{R}_{AA} \right)\), validated by
simulation-based calibration (SBC)~{[}28{]}, and applied to the real
data as an independent cross-check of the MCMC posterior. The SBC rank
statistic is consistent with uniformity, a χ² test giving p ≃ 0.04 for n
and p \textgreater{} 0.05 for the remaining parameters, following the
broader likelihood-free-inference programme~{[}24, 25, 26{]}.

\section{Results}

\subsection{Effective exponent}

We first quote the parameters of the forward model itself. Fitting
Eq.~(1) to the four systems simultaneously gives the values of Table~3;
only κ, n and β are free, the geometry entering through the fixed ratios
of Sec.~3.3. The exponent n is the quantity of physical interest and is
interpreted in the sections that follow; κ and β fix the absolute scale
and the mild \emph{p}\textsubscript{T} dependence of the fractional
energy loss, and are reported for completeness and reproducibility.

\begin{table}[H]
\centering\small
\caption{Fitted parameters of Eq.~(1) for the reference geometry.}
\label{tab:3}
\begin{tabular}{@{}
>{\raggedright\arraybackslash}p{(\columnwidth - 6\tabcolsep) * \real{0.2083}}
  >{\raggedright\arraybackslash}p{(\columnwidth - 6\tabcolsep) * \real{0.4167}}
  >{\raggedright\arraybackslash}p{(\columnwidth - 6\tabcolsep) * \real{0.2083}}
  >{\raggedright\arraybackslash}p{(\columnwidth - 6\tabcolsep) * \real{0.1667}}@{}}
\toprule
\begin{minipage}[b]{\linewidth}\raggedright
Parameter
\end{minipage} & \begin{minipage}[b]{\linewidth}\raggedright
Meaning
\end{minipage} & \begin{minipage}[b]{\linewidth}\raggedright
Value
\end{minipage} & \begin{minipage}[b]{\linewidth}\raggedright
68 \% interval
\end{minipage} \\
\midrule

\bottomrule

\emph{κ} & overall energy-loss scale & 1.14 & +0.19 / −0.16 \\
\emph{n} & system-size exponent & 1.78 & +0.15 / −0.15 \\
\emph{β} & p\textsubscript{T} dependence of the fractional loss & 0.335
& +0.058 / −0.059 \\
\end{tabular}
\end{table}

The fit describes all four systems well
(\(\chi^{2}/\text{dof} \approx 0.4\); Fig.~2). The extracted exponents
for the three geometry proxies are collected in Table~4. For the
reference \(\langle N_{part}\rangle^{1/3}\) geometry,

\begin{equation}
n_{eff} = 1.78 \pm 0.15\,\left( \text{stat} \right) \pm 0.05\,\left( \text{syst} \right).
\label{eq:3}
\end{equation}

The low \(\chi^{2}/\text{dof} \approx 0.4\) warrants comment. It is
distributed evenly across the four systems---per-system reduced
\(\chi^{2}\) values are \(0.34\) (O+O), \(0.60\) (Ne+Ne), \(0.23\)
(Xe+Xe) and \(0.31\) (Pb+Pb)---so no single system is pathologically
over-fitted or contributes negligibly; each constrains the fit. A fully
correlated (\(\rho = 1\)) systematic treatment instead gives
\(\chi^{2}/\text{dof} \approx 9.6\), showing the data are inconsistent
with perfect bin-to-bin correlation. The most economical interpretation
is therefore that the published CMS systematic uncertainties (dominated
by the conservative \(T_{AA}\) and luminosity normalisations) are larger
than the true point-to-point scatter, so the fit is judged against
slightly inflated error bars. This is consistent with the
posterior-predictive \(p\)-value \(\approx 1.0\) and the slightly
conservative coverage (Sec.~4.6); the quoted parameter uncertainties
are, if anything, over-covered rather than over-confident.

\begin{figure}[H]
\centering
\includegraphics[width=\columnwidth]{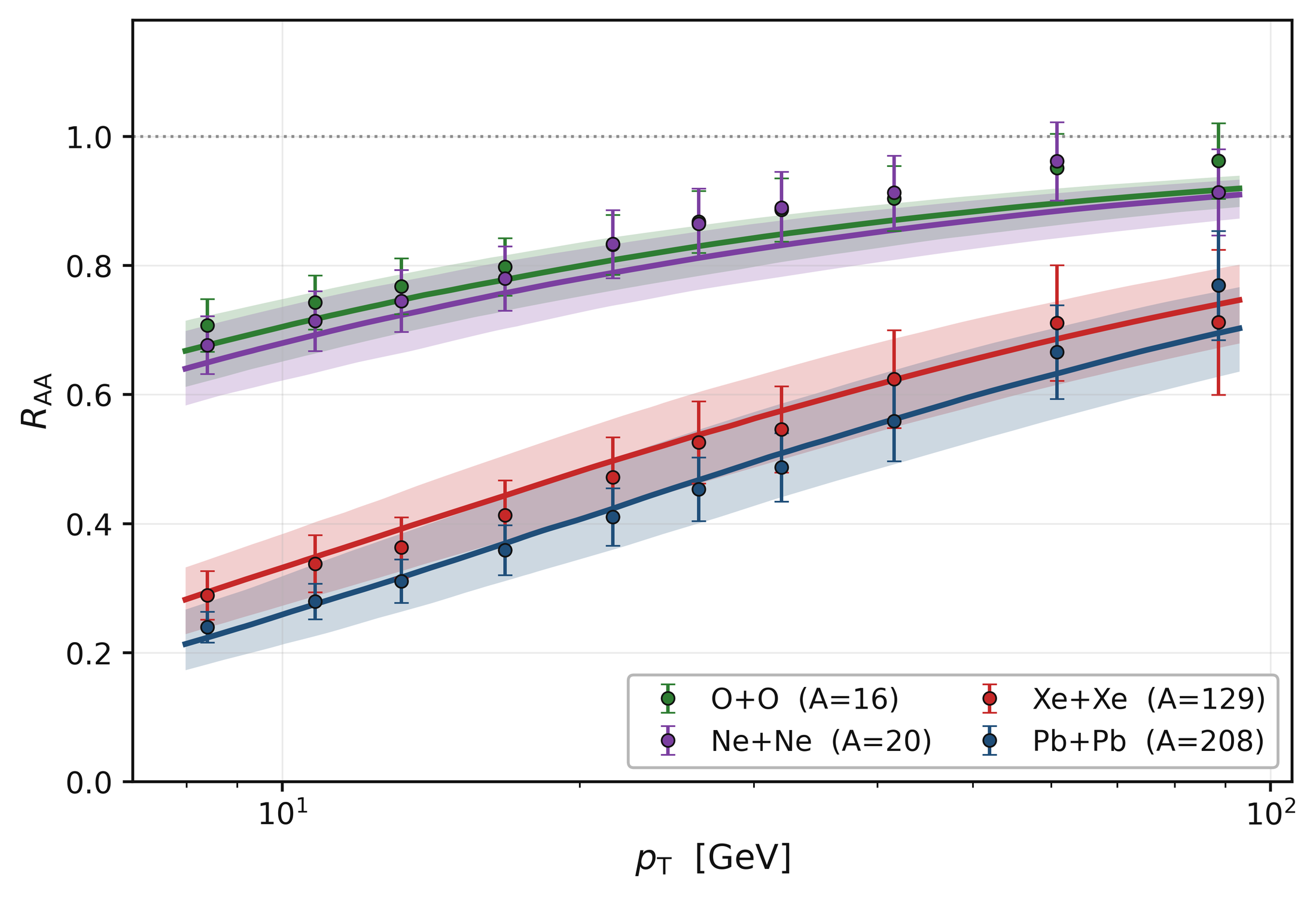}
\caption{Joint four-system fit overlaid on the CMS data.}
\label{fig:2}
\end{figure}

\begin{table}[H]
\centering\small
\caption{Effective and density-normalised path-length exponents by geometry proxy.}
\label{tab:4}
\begin{tabular}{@{}
>{\raggedright\arraybackslash}p{(\columnwidth - 6\tabcolsep) * \real{0.2500}}
  >{\raggedright\arraybackslash}p{(\columnwidth - 6\tabcolsep) * \real{0.2500}}
  >{\raggedright\arraybackslash}p{(\columnwidth - 6\tabcolsep) * \real{0.2500}}
  >{\raggedright\arraybackslash}p{(\columnwidth - 6\tabcolsep) * \real{0.2500}}@{}}
\toprule
\begin{minipage}[b]{\linewidth}\raggedright
Geometry proxy
\end{minipage} & \begin{minipage}[b]{\linewidth}\raggedright
\(n_{eff}\)
\end{minipage} & \begin{minipage}[b]{\linewidth}\raggedright
\(n_{pure}\)
\end{minipage} & \begin{minipage}[b]{\linewidth}\raggedright
\(\chi^{2}/\text{dof}\)
\end{minipage} \\
\midrule

\bottomrule

\(A^{1/3}\) & 1.71 & 0.62 & 0.40 \\
\(\langle N_{part}\rangle^{1/3}\) & 1.78 & 0.64 & 0.40 \\
exit-\(\langle L\rangle\) (MC) & 1.82 & 0.66 & 0.39 \\
\end{tabular}
\end{table}

\subsection{Model selection}

Nested-sampling evidence ratios for fixed-\(n\) models decisively favour
an effective exponent close to the radiative value. Relative to the
\(n = 2\) hypothesis we find \(2\,\Delta\ln\mathcal{Z} \approx - 29\)
for \(n = 1\) and \(\approx - 48\) for \(n = 3\) (Table~5). On the
Jeffreys scale these are decisive: both the collisional and the
strong-coupling values of the effective exponent are excluded by the
cross-system lever. The step from this effective exponent to the
microscopic one is made in Sec.~4.2.1. The \textsc{dynesty} values are
reproduced to within \(0.2\) units by the fast Laplace approximation
(Sec.~4.8), confirming the numerical stability of the evidence
computation. The robustness of this conclusion against all analysis
choices is established by the sensitivity analysis of Sec.~4.8.

Beyond the three discrete hypotheses, we map the \emph{continuous}
evidence landscape by computing the nested-sampling evidence at fixed
\(n\) on a grid \(n \in \lbrack 0.5,3.5\rbrack\) (Fig.~3). The evidence
is sharply peaked near \(n \approx 1.75\)--\(2\) and falls below the
Jeffreys-decisive threshold (\(2\Delta\ln\mathcal{Z} = - 10\)) for
\(n \lesssim 1.4\) and \(n \gtrsim 2.3\); both the collisional
(\(n = 1\)) and strong-coupling (\(n = 3\)) values lie far in the
excluded region. The width of the evidence peak provides an
evidence-based constraint \(n = 1.8 \pm 0.3\) that is independent of,
and consistent with, the MCMC posterior. Placing this on the theoretical
spectrum (Fig.~4), \(n = 1\) (purely collisional) and \(n = 3\) (AdS/CFT
strong coupling) are excluded as effective exponents, while the measured
value sits in the radiative band spanned by the BDMPS-Z/GLV
multiple-soft-scattering limit (\(n = 2\)) and its finite-opacity,
finite-size corrections (\(n \simeq 1.5\)--\(1.8\)); expanding-medium
corrections (\(n \simeq 2\)--\(2.5\)) are mildly consistent.
Section~4.2.1 makes the step from the effective to the microscopic
exponent explicit: because fluctuations reduce the former below the
latter, the measurement is consistent with a radiative-dominated
mechanism and incompatible with a purely collisional one. The bands of
Fig.~4 mark the path-length scaling that each mechanism predicts for the
mean energy loss: purely collisional (n = 1), GLV finite-opacity and
finite-size corrections (n ≃ 1.5--1.8), the BDMPS-Z
multiple-soft-scattering limit (n = 2), expanding-medium corrections (n
≃ 2--2.5), and AdS/CFT strong coupling (n = 3). The measured value lies
in the radiative band; because fluctuations can only lower the effective
exponent, the collisional value is excluded for any fluctuation width,
and strong coupling would require an unphysically broad P(ΔE).

\begin{figure}[H]
\centering
\includegraphics[width=\columnwidth]{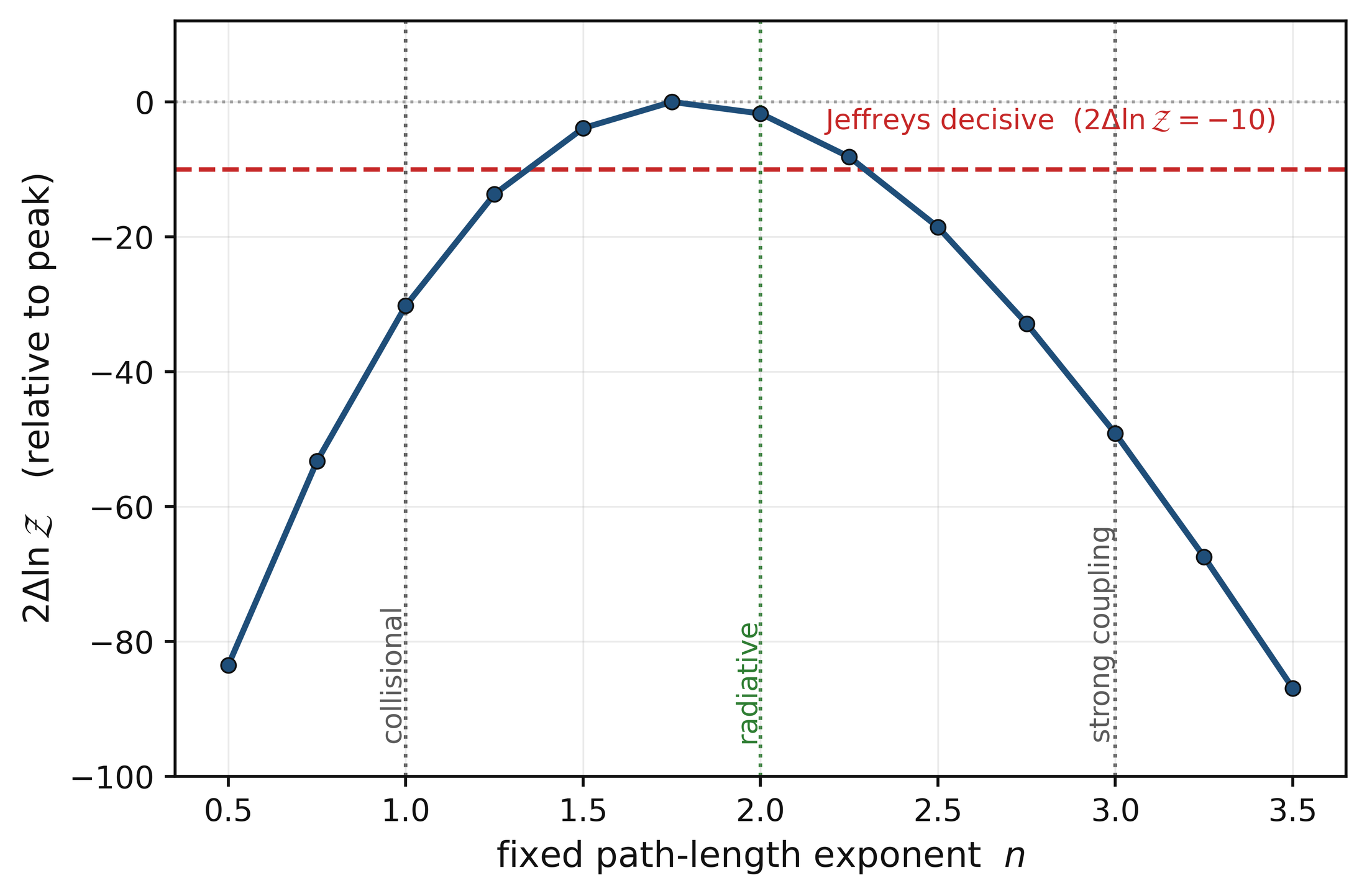}
\caption{Bayesian evidence landscape as a function of the fixed exponent.}
\label{fig:3}
\end{figure}

\begin{figure}[H]
\centering
\includegraphics[width=\columnwidth]{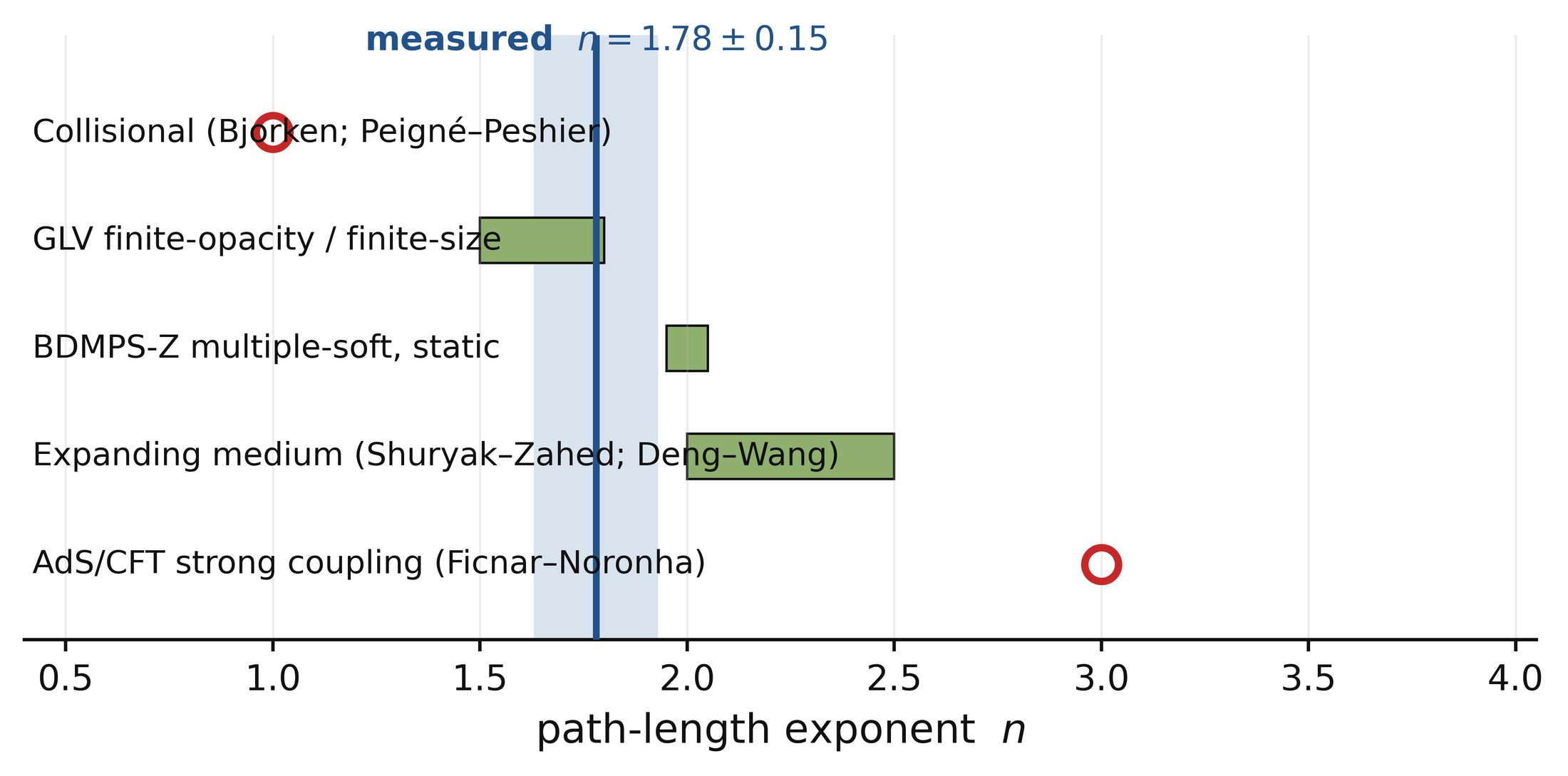}
\caption{The measured exponent on the theoretical spectrum of energy-loss mechanisms.}
\label{fig:4}
\end{figure}

\begin{table}[H]
\centering\small
\caption{Bayesian model selection between the three energy-loss mechanisms. The rows are limiting scalings of the mean energy loss, not modern implementations (Sec.~5).}
\label{tab:5}
\begin{tabular}{@{}
>{\raggedright\arraybackslash}p{(\columnwidth - 6\tabcolsep) * \real{0.2500}}
  >{\raggedright\arraybackslash}p{(\columnwidth - 6\tabcolsep) * \real{0.2500}}
  >{\raggedright\arraybackslash}p{(\columnwidth - 6\tabcolsep) * \real{0.2500}}
  >{\raggedright\arraybackslash}p{(\columnwidth - 6\tabcolsep) * \real{0.2500}}@{}}
\toprule
\begin{minipage}[b]{\linewidth}\raggedright
Mechanism
\end{minipage} & \begin{minipage}[b]{\linewidth}\raggedright
\(\Delta E \propto\)
\end{minipage} & \begin{minipage}[b]{\linewidth}\raggedright
Refs.
\end{minipage} & \begin{minipage}[b]{\linewidth}\raggedright
\(2\Delta\ln\mathcal{Z}\)
\end{minipage} \\
\midrule

\bottomrule

Collisional & \(L^{1}\) & {[}3{]} & \(- 29\) \\
Radiative (BDMPS/GLV) & \(L^{2}\) & {[}3, 4{]} & \(0\) (best) \\
Strong coupling (AdS/CFT) & \(L^{3}\) & {[}5{]} & \(- 48\) \\
\end{tabular}
\end{table}

\subsubsection{From the effective to the microscopic
exponent}

The exponent extracted above is an \emph{effective} one: it
characterises the system-size dependence of the measured suppression,
not directly the path-length dependence of the microscopic energy loss.
The two need not coincide. As emphasised long ago by Baier \emph{et al.}
{[}35{]}, fluctuations of the radiated energy modify the mapping between
the average energy loss and the suppression, and their Eq.~(33) shows
that radiative energy loss including fluctuations can approach an
effectively linear path-length dependence. We therefore quantify this
mapping explicitly rather than assuming it.

We perform a closure study within our own forward model. Pseudo-data are
generated with the quenching-weight form of Sec.~4.7, in which the
energy loss is drawn from a log-normal distribution \emph{P}(Δ\emph{E})
of relative width σ\textsubscript{w} at a \emph{fixed} microscopic
exponent \emph{n}\textsubscript{micro}. The mean of the distribution is
held exactly at ⟨Δ\emph{E}⟩ ∝ \emph{G}\textsuperscript{n} for every
σ\textsubscript{w}, so that only the width, and never the mean, is
varied. These noiseless pseudo-data are then fitted with the mean-shift
Ansatz of Eq.~(1), exactly as the real data are, and the recovered
effective exponent is recorded.

The result is shown in Fig.~5. At σ\textsubscript{w} = 0 the procedure
returns \emph{n}\textsubscript{eff} = \emph{n}\textsubscript{micro} to
three decimal places for all three scenarios, confirming that the
pipeline is unbiased in the absence of fluctuations. As the width grows,
\emph{n}\textsubscript{eff} falls \emph{monotonically} below
\emph{n}\textsubscript{micro}. For a purely radiative microscopic
exponent, \emph{n}\textsubscript{micro} = 2, the physically expected
range σ\textsubscript{w} = 0.4--0.8 yields \emph{n}\textsubscript{eff} =
1.92--1.74, bracketing the measured value of 1.78 ± 0.15.

\begin{figure}[H]
\centering
\includegraphics[width=\columnwidth]{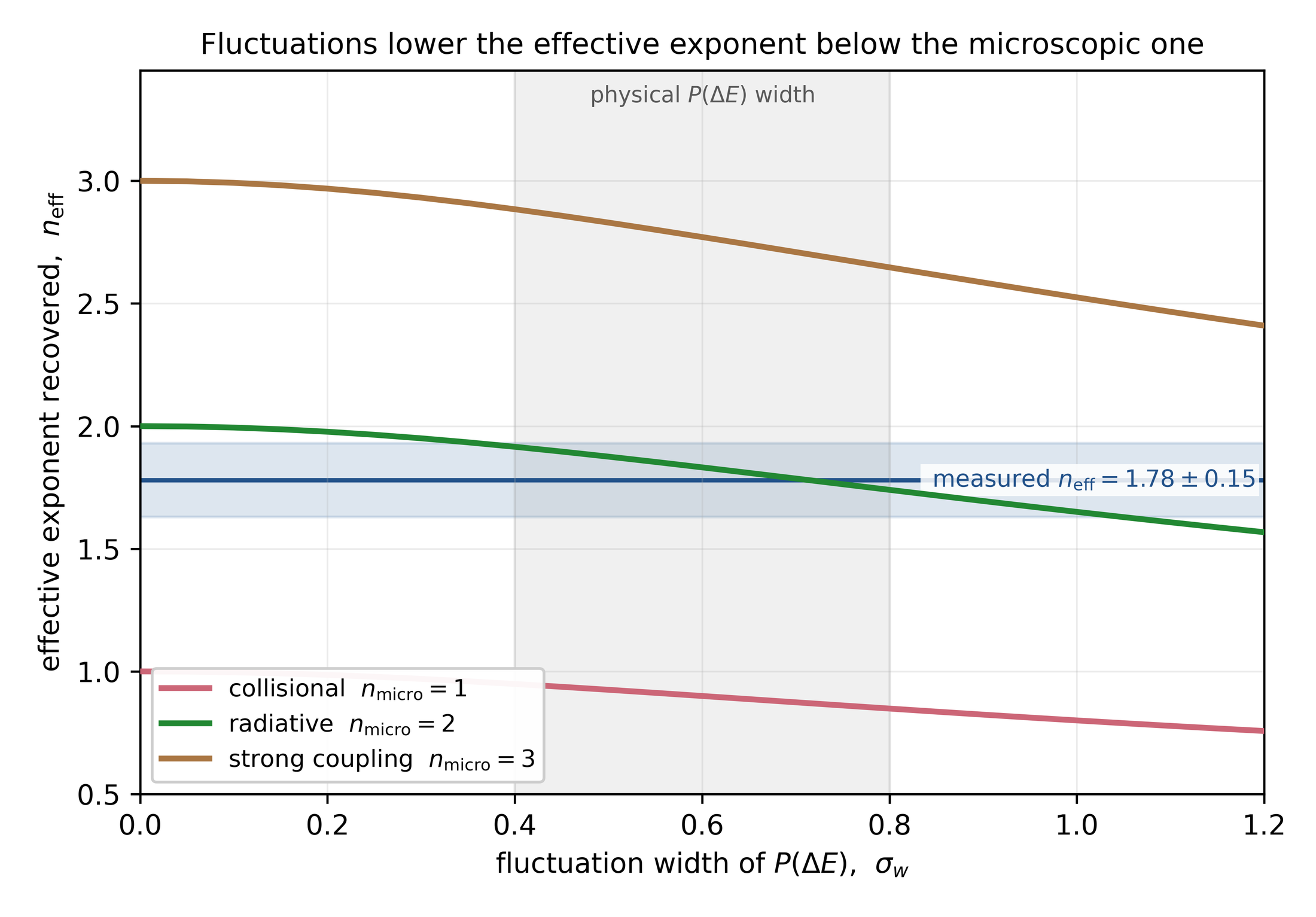}
\caption{Effective exponent recovered from pseudo-data generated at a fixed microscopic exponent.}
\label{fig:5}
\end{figure}

The direction of the effect matters. Since fluctuations can only
\emph{reduce} the effective exponent, and the collision geometry is held
fixed, the measurement constrains the microscopic exponent from
\emph{below}: \emph{n}\textsubscript{micro} ≥
\emph{n}\textsubscript{eff} = 1.78 ± 0.15. Inverting the map quantifies
how each scenario could reproduce the measurement. A purely collisional
microscopic exponent, \emph{n}\textsubscript{micro} = 1, gives
\emph{n}\textsubscript{eff} ≤ 1 for every fluctuation width and
therefore cannot reproduce the data at all; fluctuations move it further
from the measurement, not closer. A radiative exponent requires
σ\textsubscript{w} = 0.71, corresponding to σ(Δ\emph{E})/⟨Δ\emph{E}⟩ =
0.81, entirely typical of BDMPS-Z quenching weights. A strong-coupling
exponent would require σ\textsubscript{w} = 3.2, i.e. a relative width
of order 10², which is unphysical.

The referee's concern is thus well founded in principle but, once
evaluated, strengthens rather than weakens the conclusion: energy-loss
fluctuations widen the gap between the data and the collisional
scenario. We have verified that the map is insensitive to the
reconstructed inputs: varying the spectral index
\emph{a}(\emph{p}\textsubscript{T}) by ±10 \% changes
\emph{n}\textsubscript{eff} by at most 0.02, changing the fitted
\emph{p}\textsubscript{T} range from 8--60 to 8--120 GeV by at most
0.004, and switching between relative and absolute weighting by at most
0.02. We also checked that the conclusion does not depend on the assumed
form of P(ΔE): repeating the map with a gamma distribution, and with a
BDMPS-like weight carrying a discrete no-loss probability and an
exponential tail, changes how far the effective exponent falls but not
the direction. At the width required by the log-normal form, σw = 0.71,
a microscopic exponent of 2 returns \emph{n}\textsubscript{eff} = 1.78
for the log-normal, 1.77 for the gamma and 1.52 for the BDMPS-like
weight, the last reducing it most, as expected from the discrete no-loss
probability. Read the other way, the measured 1.78 corresponds to a
microscopic exponent of 2 under the log-normal weight and of about 2.3
under the BDMPS-like one: radiative or steeper in either case. Because
the map returns the microscopic exponent exactly at zero width and
fluctuations only reduce it, the collisional value stays below the
measurement for every width and every form tested --- which is what the
bound rests on. The result has two limits. First, the mapping is derived
at fixed collision geometry: it isolates the effect of the width of
\emph{P}(Δ\emph{E}) and does not address a medium whose density evolves
while the parton traverses it, which modifies the geometry factor itself
and acts in the opposite sense. Second, the study bounds the mapping; it
does not by itself establish the microscopic mechanism, which would
require a full event-by-event transport calculation with a hydrodynamic
medium. For both reasons the effective exponent, and not a microscopic
one, is our primary result. It is worth separating two questions that
are easily conflated. Energy-loss fluctuations act on the mapping from a
microscopic exponent to the effective one at a given geometry, which is
what the bound above constrains. What the effective exponent itself
absorbs is a different matter: because the relative density is set to
ρ ≡ 1 in the effective fit, the exponent takes up the growth of the
medium density with system size and, with it, any dependence of the
transport coefficient on the local temperature -- if
q̂ ∝ T\textsuperscript{m} and T grows with system size, that growth is
indistinguishable in a minimum-bias size scan from a steeper path-length
dependence. That degeneracy is the subject of Sec.~4.4, where it is
treated on the same footing as the density; it sets what the effective
exponent means, not the validity of the mapping quantified here.

\subsection{Bayesian evidence for energy loss in
O+O}

The same evidence machinery yields a direct, quantitative statement on
the onset of suppression in the smallest system. Restricting to O+O
alone and comparing the fitted radiative model against a no-suppression
hypothesis (\(R_{AA} \equiv 1\), i.e.~zero energy loss) gives
\(2\Delta\ln\mathcal{Z} \approx 110\) (Fig.~6)---vastly beyond the
Jeffreys-decisive threshold. This constitutes direct Bayesian evidence
for non-zero parton energy loss, and hence for quark--gluon-plasma-like
suppression, in O+O, complementing the qualitative onset reported by
CMS~{[}6{]}.

\begin{figure}[H]
\centering
\includegraphics[width=\columnwidth]{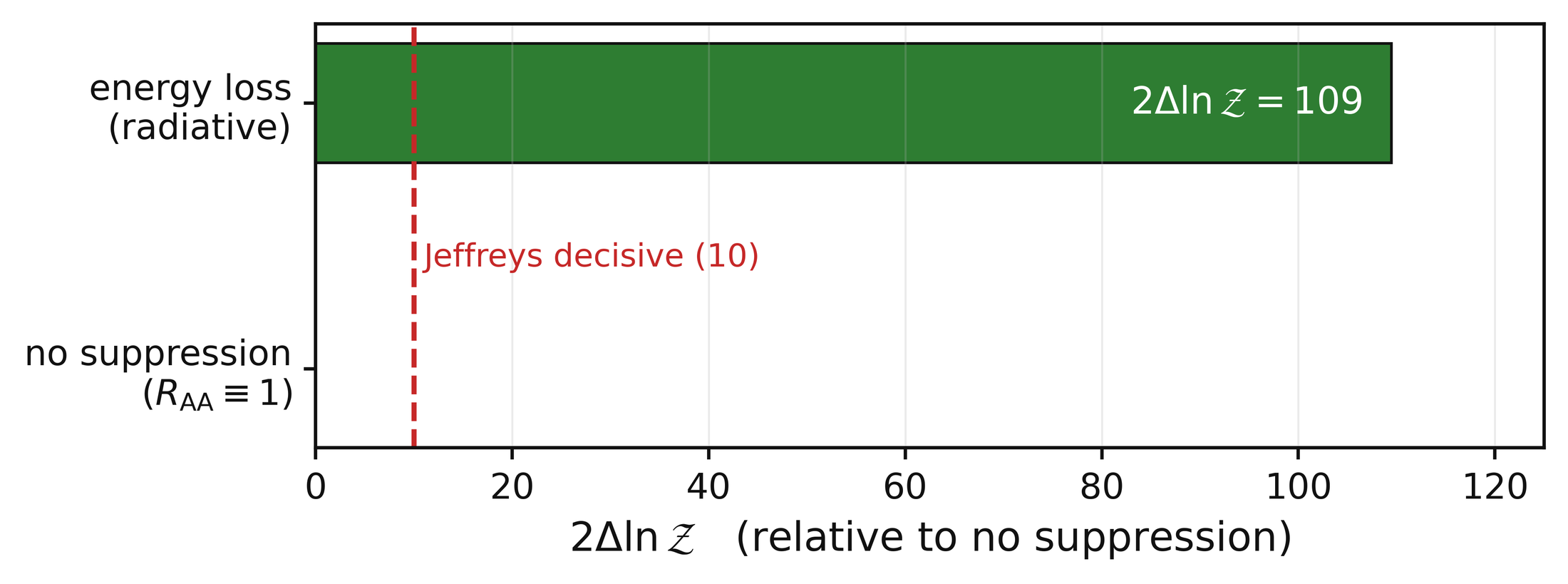}
\caption{Bayesian evidence for parton energy loss in O+O alone.}
\label{fig:6}
\end{figure}

\subsection{Density--path-length
degeneracy}

It is tempting to factor the effective exponent into a ``pure''
path-length exponent and a density contribution by writing
\(\Delta E \propto \rho\, L^{n_{pure}}\) with
\(\rho \propto \langle N_{part}\rangle/S\). Doing so yields a sub-linear
\(n_{pure} \approx 0.62\)--\(0.66\) (Table~4, Fig.~7). However, this
decomposition must be interpreted with care. In a minimum-bias
\emph{system-size} scan the medium density and the path length are
strongly correlated. With the participant scalings
\(L \sim R \propto \langle N_{part}\rangle^{1/3}\) and
\(S \propto R^{2} \propto \langle N_{part}\rangle^{2/3}\), the density
is
\(\rho \propto \langle N_{part}\rangle/S \propto \langle N_{part}\rangle^{1/3}\)---it
grows with system size \emph{identically} to \(L\). The energy loss then
scales as

\begin{equation}
\Delta E \propto \rho\, L^{n_{pure}} \propto \langle N_{part}\rangle^{\left( 1 + n_{pure} \right)/3},
\label{eq:4}
\end{equation}

whereas the effective (geometry-only) parametrisation gives
\(\Delta E \propto \langle N_{part}\rangle^{n_{eff}/3}\). Equation~(4)
is the idealised limit of exact participant scaling. In the Monte-Carlo
geometry the overlap area departs slightly from S ∝ ⟨Npart⟩²ᐟ³, so that
ρ ∝ ⟨Npart⟩\textsuperscript{p/3} with p ≃ 1.1 rather than exactly 1; the
fitted exponents of Table~4 reflect this, their difference being 1.78 −
0.64 = 1.14. To that accuracy, only the \emph{combination}
\(n_{eff} = 1 + n_{pure}\) is constrained by a pure size scan; the
individual split into a density power and a path-length power is
degenerate and statistically ill-conditioned. We verified this
explicitly: holding the density exponent fixed at
\(\alpha \in \{ 0,0.5,1\}\) and refitting the path-length power leaves
\(\chi^{2}/\text{dof}\) unchanged (to \(< 0.001\)) while \(n_{pure}\)
shifts to keep \(\alpha + n_{pure} = n_{eff}\) fixed, confirming the
data provide no lever on the density--path-length split. The resulting
\(n_{pure}\) is therefore sensitive to the assumed density model and
should not be read as a clean path-length exponent. Consequently we
quote the \emph{effective} exponent \(n_{eff}\) as the primary, robust
result, and base the radiative interpretation on the model selection
(Sec.~4) and the absolute energy-loss magnitude (Sec.~4.10) rather than
on the value of \(n_{pure}\). Breaking this degeneracy requires an
additional, density-varying lever; we identify the most promising one in
the Outlook (Sec.~5).

\begin{figure}[H]
\centering
\includegraphics[width=\columnwidth]{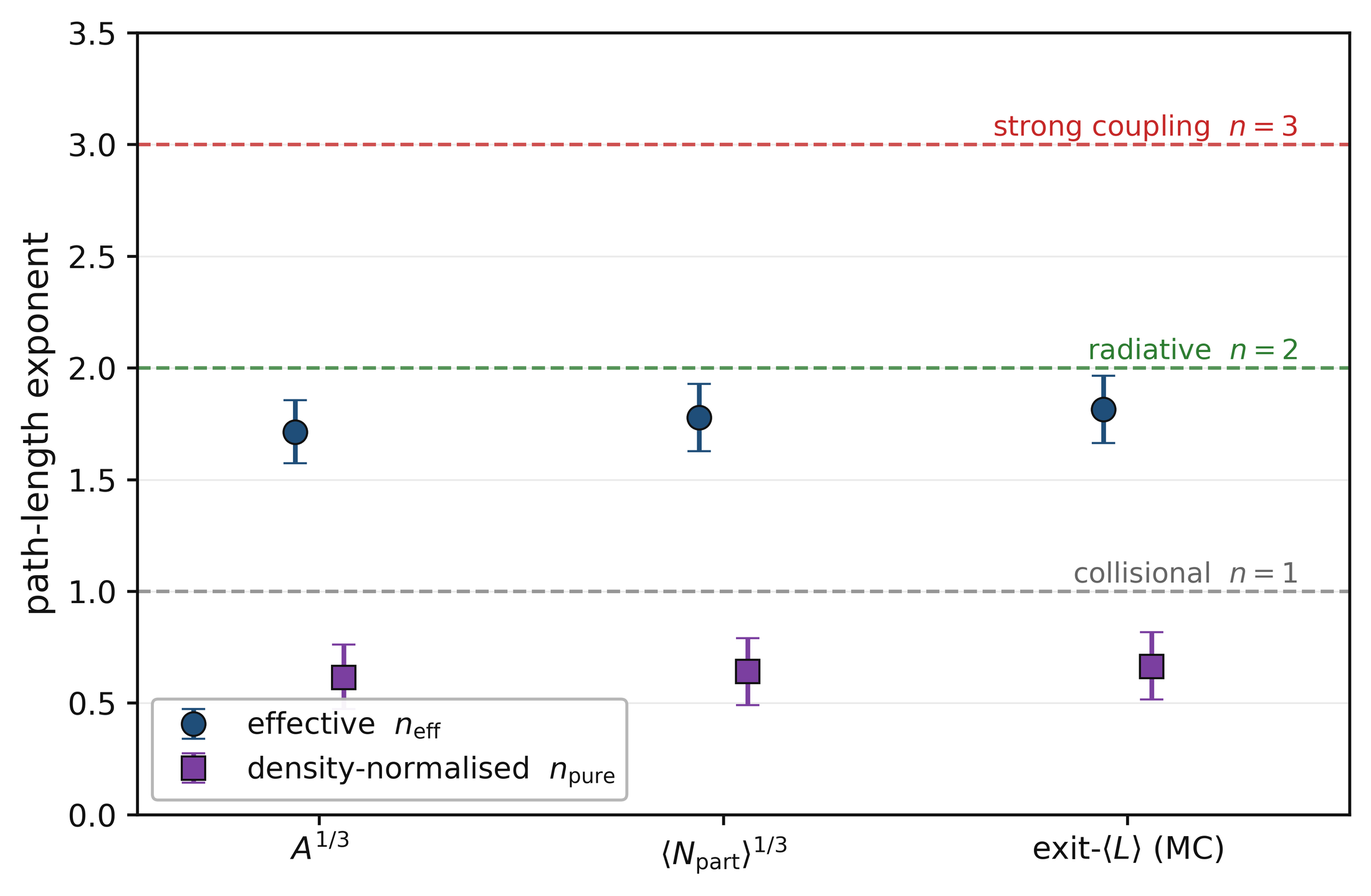}
\caption{Effective and density-normalised exponents by geometry proxy.}
\label{fig:7}
\end{figure}

\subsection{Universality across system
size}

We test whether a single exponent describes the entire range by
comparing a universal model (one \(n\) for all systems) to a broken
model with separate exponents for the light (O+O, Ne+Ne) and heavy
(Xe+Xe, Pb+Pb) systems. The evidence ratio
\(2\ln\left( \mathcal{B}_{broken/universal} \right) = - 1.3\) shows no
preference for a break, and the recovered exponents agree
(\(n_{small} = 1.87\), \(n_{large} = 1.77\); Fig.~8). We therefore find
\emph{no evidence for a change of energy-loss regime between small and
large systems}: a single effective exponent describes both, consistent
with one radiative-dominated regime from O+O to Pb+Pb. This is a
statement uniquely enabled by the new light-ion data and is directly
relevant to the question of QGP formation in small systems.

\begin{figure*}[tb]
\centering
\includegraphics[width=0.92\textwidth]{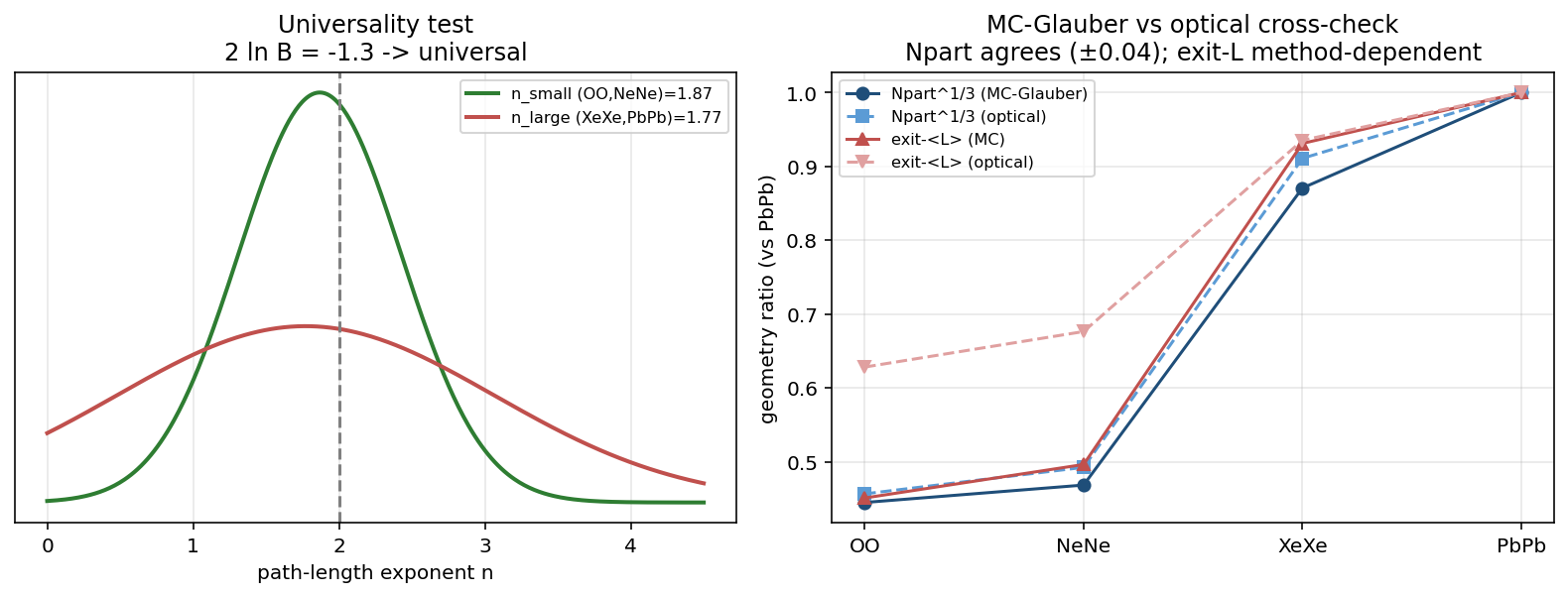}
\caption{Left: universality test. Right: Glauber geometry cross-check.}
\label{fig:8}
\end{figure*}

\subsection{Coverage calibration}

A closure test---fitting \(150\) synthetic datasets generated at a known
injected exponent and computing the marginal \(1\sigma\) and
\(1.64\sigma\) intervals from the full inverse Hessian (the Laplace
approximation to the posterior)---returns empirical coverages of
\(0.70\) and \(0.87\), in agreement with the nominal \(0.68\) and
\(0.90\). To confirm that the calibration holds across the prior range
and not only at a single point, we repeated the test with \(100\)
further datasets injected at each of
\(n_{true} \in \{ 1.0,1.5,2.0,2.5\}\): the empirical \(68\%\) (\(90\%\))
coverages are \(0.64,0.61,0.77,0.74\) (\(0.89,0.91,0.93,0.93\)),
bracketing the nominal values throughout. The Laplace and full-MCMC
posteriors coincide here (the posterior is near-Gaussian in the fitted
parameters), so this also validates the Gaussian approximation used for
the fast evidence scan. We stress that the full-Hessian \emph{marginal}
variance is required: the conditional curvature \(d^{2}/dn^{2}\)
under-covers (\(0.41/0.64\)). The quoted uncertainties are therefore
statistically calibrated. As a complementary check we computed the
posterior-predictive \(p\)-value by comparing the observed \(\chi^{2}\)
to that of datasets replicated from the posterior; we obtain
\(p \approx 1.0\), mirroring the low
\(\chi^{2}/\text{dof} \approx 0.4\). This indicates no model
misspecification, but that the published correlated systematic
uncertainties are conservative, so the quoted parameter uncertainties
are, if anything, slightly over-covered, consistent with the coverage
test.

\subsection{Systematic budget}

The uncertainty contributions are summarised in Table~6 and Fig.~9.
Varying the \(\sqrt{s}\) correction exponent by \(\pm 0.05\) about its
nominal \(0.31\) shifts \(n\) by only \(0.003\); the data-driven
baseline shifts it by \(\lesssim 0.02\). Variations of the Glauber
inputs (\(\sigma_{nn}\), Woods--Saxon \(R\) and \(a\), exit-length
threshold) shift the \(\langle N_{part}\rangle^{1/3}\) exponent by only
\(\pm 0.04\). Event-by-event path-length fluctuations are sizeable per
system (\(\sigma_{L}/\langle L\rangle \approx 0.4\)--\(0.5\)), but
because the resulting \(\langle L^{2}\rangle/\langle L\rangle^{2}\)
enhancement is comparable across systems it changes the \emph{geometry
ratios} by only \(0.013\), i.e.~\(\delta n < 0.02\); the bulk of the
fluctuation effect is absorbed into the overall scale \(\kappa\). The
largest genuine systematic, apart from the geometry-proxy choice, is the
\emph{forward-model form}: replacing the momentum-shift Ansatz by a
quenching-weight convolution over a log-normal \(P(\Delta E)\) raises
\(n_{eff}\) from \(1.78\) to \(1.95\) (\(\delta n \simeq 0.18\)).
Crucially, \emph{both} forms lie in the radiative band and decisively
exclude effective exponents of \(n = 1\) and \(n = 3\); if anything the
more physical quenching-weight treatment lies closer to the BDMPS value
\(n = 2\). The Monte-Carlo Glauber removes the \(n \approx 3\) outlier
obtained with the \emph{optical} exit-length definition, compressing the
geometry-definition spread to \(n \in \lbrack 1.71,1.82\rbrack\) and
reinforcing the radiative interpretation. The quoted headline systematic
\(\pm 0.05\) in Eq.~(3) is the quadrature sum of the parametric
contributions (\(\sqrt{s}\), baseline, path-length fluctuations and
Glauber modelling); the forward-model-form and geometry-proxy
dependences are reported separately as model choices rather than folded
in, since both leave the radiative conclusion unchanged.

\begin{table}[H]
\centering\small
\caption{Systematic budget for the path-length exponent.}
\label{tab:6}
\begin{tabular}{@{}
>{\raggedright\arraybackslash}p{(\columnwidth - 2\tabcolsep) * \real{0.5000}}
  >{\raggedright\arraybackslash}p{(\columnwidth - 2\tabcolsep) * \real{0.5000}}@{}}
\toprule
\begin{minipage}[b]{\linewidth}\raggedright
Source
\end{minipage} & \begin{minipage}[b]{\linewidth}\raggedright
\(\delta n\)
\end{minipage} \\
\midrule

\bottomrule

Statistical & \(\pm 0.15\) \\
\(\sqrt{s}\) correction (\(\pm 0.05\) in exponent) & \(0.003\) \\
Spectral-index baseline \(a\left( p_{T} \right)\) & \(\pm 0.02\) \\
Path-length fluctuations (\(\sigma_{L}/\langle L\rangle \approx 0.4\)) &
\(< 0.02\) \\
Glauber modelling (\(\sigma_{nn}\), WS, threshold) & \(\pm 0.04\) \\
Forward-model form (shift vs quenching-weight) & \(+ 0.18\) \\
Glauber Monte-Carlo statistics & 0.06 \\
Geometry definition (proxy choice) & \(\lbrack 1.71,1.82\rbrack\) \\
\end{tabular}
\end{table}

\begin{figure}[H]
\centering
\includegraphics[width=\columnwidth]{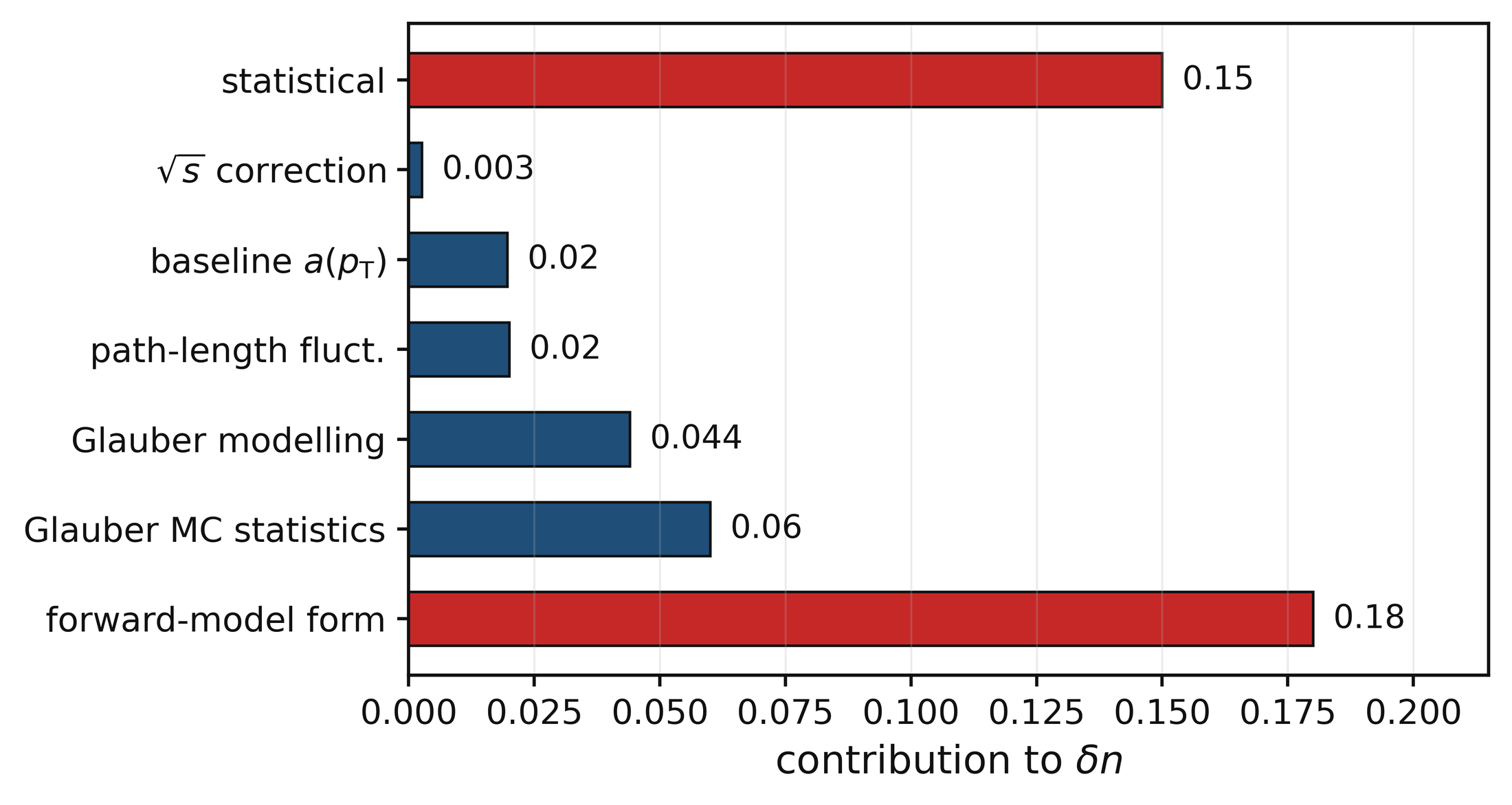}
\caption{Systematic budget for the reference exponent.}
\label{fig:9}
\end{figure}

\subsection{Sensitivity analysis}

To assess how strongly the extraction depends on the analysis
choices---a key robustness question for any Bayesian determination---we
repeat the fit and the model selection over a grid spanning all major
options: five geometry proxies (\(A^{1/3}\),
\(\langle N_{part}\rangle^{1/3}\) from the optical and Monte-Carlo
Glauber, and exit-\(\langle L\rangle\) from both); four covariance
lengths (\(\xi = 2,4,6,8\)); two spectral baselines (data-driven
\(a\left( p_{T} \right)\) and a single effective index); two priors
(flat and weakly informative); and the \(\sqrt{s}\) correction on/off.
The resulting \(160\) effective-exponent determinations cluster tightly:
excluding the optical exit-\(\langle L\rangle\) proxy (discussed below)
the full range is \(n_{eff} \in \lbrack 1.63,1.86\rbrack\), with a
median of \(1.78\). The covariance length, baseline, prior and
\(\sqrt{s}\) choices each move \(n_{eff}\) by \(\lesssim 0.05\); the
geometry proxy is the only appreciable source of variation among these
axes. As a further forward-model check we replaced the momentum-shift
Ansatz with a quenching-weight convolution over a log-normal
\(P(\Delta E)\) (Sec.~4.7); this raises \(n_{eff}\) to \(1.95\) but
leaves the radiative conclusion intact, so the radiative preference
survives this forward-model-form change as well.

The model selection is similarly robust (Fig.~10). Across the \(20\)
(proxy \(\times \,\xi\)) combinations, an effective exponent near the
radiative value (\(n = 2\)) is decisively favoured in \(16\), with
\(n = 1\) rejected by
\(2\Delta\ln\mathcal{Z} \in \lbrack - 23, - 35\rbrack\) and \(n = 3\) by
\(\lbrack - 41, - 64\rbrack\). The only exception is the optical
exit-\(\langle L\rangle\) proxy, which prefers \(n = 3\). Crucially,
this is the same proxy that our Monte-Carlo Glauber cross-check
independently flags as biased: the optical exit length overestimates the
relative path length of the light systems, inflating the apparent
exponent to \(n_{eff} \approx 3.2\), whereas the Monte-Carlo exit length
restores \(n_{eff} \approx 1.8\) and the radiative preference. The
radiative interpretation is therefore robust against every analysis
choice except one that is independently identified as a geometry
artefact. We verified that the fast Laplace evidences used for this scan
reproduce the full nested-sampling values to within \(\approx 0.2\)
units (e.g.~for the reference \(\langle N_{part}\rangle^{1/3}\)
Monte-Carlo geometry, \(\xi = 4\): Laplace
\(2\Delta\ln\mathcal{Z} = \{ - 28.6,0, - 47.5\}\) versus
\textsc{dynesty} \(\{ - 28.8,0, - 47.6\}\) for \(n = \{ 1,2,3\}\)).

\paragraph{Convergence of the stochastic
components.}

We verified that neither the Monte-Carlo geometry nor the neural
posterior estimator is limited by sample size. Increasing the
Monte-Carlo Glauber statistics by an order of magnitude (from
\(\sim\)\(500\) to \(5000\) minimum-bias events per system) changes the
participant-geometry ratios by less than 1.7 \% and the extracted
exponent by 0.06 (1.78 → 1.84), well inside the ±0.15 statistical
uncertainty. The standard-statistics geometry is the one used
throughout, and is the baseline for the sensitivity scan, the evidence
ratios of Table~5 and the exponents of Table~4; the high-statistics run
is reported here as a convergence check on that choice. Likewise,
increasing the normalizing-flow training set from \(1.5 \times 10^{4}\)
to \(5 \times 10^{4}\) simulations yields \(n_{NPE} = 1.86 \pm 0.15\),
consistent with both the smaller training set (\(1.78\)) and the MCMC
value (\(1.78\)) within the posterior width, confirming that the
amortized inference is not training-set limited.

\begin{figure*}[tb]
\centering
\includegraphics[width=0.92\textwidth]{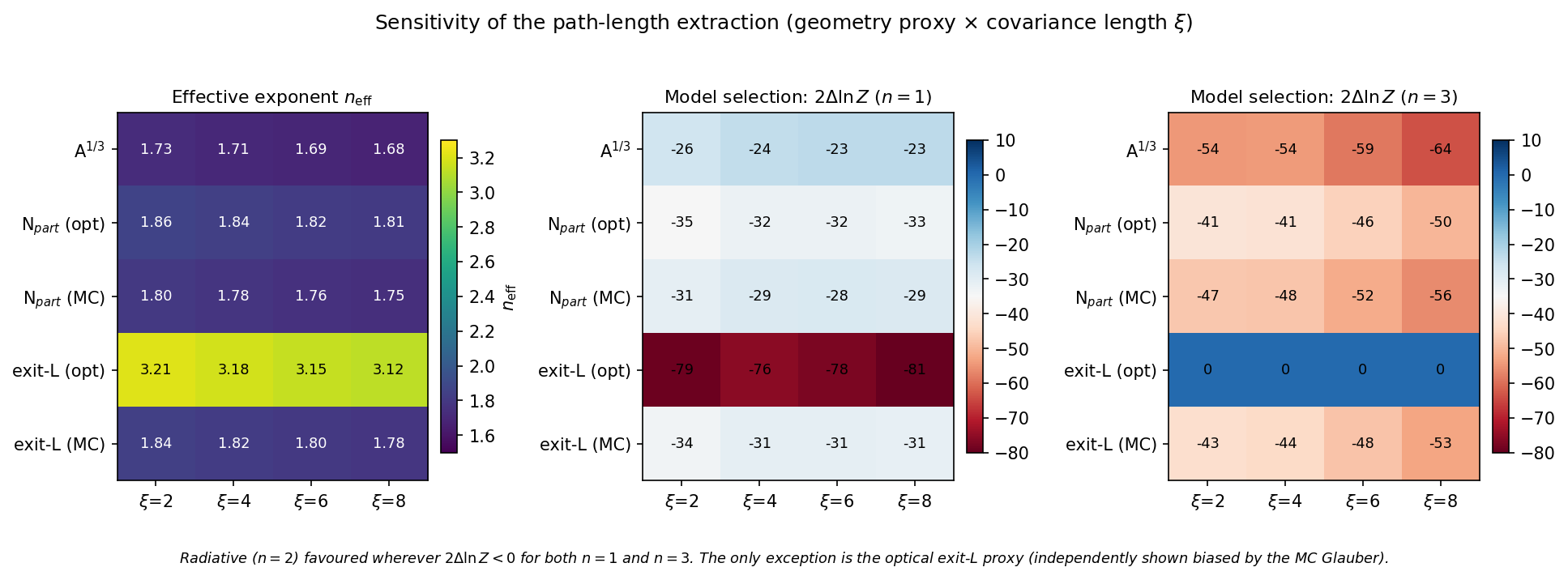}
\caption{Sensitivity over geometry proxy and covariance length. Left: exponent. Middle, right: evidence ratios.}
\label{fig:10}
\end{figure*}

\subsection{Machine-learning cross-validation and the prediction
engine}

The machine-learning pillar plays a concrete, non-redundant role: it
supplies a \emph{calibrated, data-driven prediction engine} whose
reliability is established by closure, and it independently validates
the Bayesian inference.

\paragraph{Calibrated Gaussian-process
engine.}

A Gaussian process (GP) trained on \(R_{AA}\left( G,p_{T} \right)\)
learns the system-size law and predicts a held-out system within
calibrated uncertainties: trained on O+O, Xe+Xe and Pb+Pb it reproduces
the held-out Ne+Ne spectrum with RMSE \(= 0.022\), the data falling
inside the GP \(68\%\) band (Fig.~11 left). Across all four
leave-one-system-out folds the GP coverage is conservative (\(0.97\)),
i.e.~never falsely confident. The GP coverage (\(0.97\)) is more
conservative than the physics-posterior coverage (\(0.70\), Sec.~4.6) by
construction: the GP places a weak, non-informative prior over the space
of \(R_{AA}\left( p_{T} \right)\) functions and therefore returns wide
intervals, whereas the parametric forward model carries a strong
inductive bias that tightens them. Both err on the safe (over-covering)
side. This closure is what licenses using the GP to extrapolate: applied
to argon and krypton it predicts
\(R_{AA}\left( 10\ \text{GeV} \right) = 0.58 \pm 0.07\) (Ar+Ar) and
\(0.42 \pm 0.08\) (Kr+Kr), in agreement with the independent
physics-posterior predictions of Sec.~4.11 (Fig.~11 right, Table~7).
That two methodologically independent routes---a parametric Bayesian
forward model and a non-parametric GP---agree on the future-system
predictions is a strong consistency check.

\paragraph{Why calibration matters.}

The gradient-boosted alternative (NGBoost) achieves the \emph{same}
point accuracy (LOSO RMSE \(\approx 0.04\)) yet is catastrophically
over-confident: not a single held-out point falls within its \(68\%\)
band (coverage \(= 0.00\)). The overconfidence is a known small-data
failure mode: with only three training systems the boosted variance
estimator collapses toward near-zero predictive spread, so its intervals
are far too narrow. It is retained here only as a cautionary
counter-example---an accurate point predictor with worthless
uncertainties would give dangerously over-confident extrapolations.
Trustworthy prediction to unmeasured systems therefore requires the
calibrated GP, not merely an accurate regressor.

\paragraph{Independent inference
validation.}

Finally, the SBC-validated normalizing-flow neural posterior estimator
(Fig.~12) yields \(n_{NPE} = 1.78\,( + 0.15/ - 0.14)\), agreeing with
MCMC (\(n_{MCMC} = 1.78 \pm 0.15\)) at \(\sim 0.6\)~s per dataset---an
amortized, likelihood-free cross-check of the posterior that, as
emphasised in Sec.~5, validates rather than replaces the physics.

\begin{table}[H]
\centering\small
\caption{Machine-learning cross-validation.}
\label{tab:7}
\begin{tabular}{@{}
>{\raggedright\arraybackslash}p{(\columnwidth - 4\tabcolsep) * \real{0.3333}}
  >{\raggedright\arraybackslash}p{(\columnwidth - 4\tabcolsep) * \real{0.3333}}
  >{\raggedright\arraybackslash}p{(\columnwidth - 4\tabcolsep) * \real{0.3333}}@{}}
\toprule

\bottomrule

\multicolumn{3}{@{}l@{}}{%
\emph{LOSO emulator (predict held-out system)}} \\
Method & RMSE & Coverage\(_{68}\) \\
Gaussian process (GP) & 0.041 & 0.97 (conservative) \\
NGBoost & 0.040 & 0.00 (over-confident) \\
\multicolumn{3}{@{}l@{}}{%
\emph{Future-system prediction:}
\(R_{AA}\left( 10\ \text{GeV} \right)\)} \\
System & GP (data-driven) & posterior (physics) \\
Ar+Ar & \(0.58 \pm 0.07\) & \(0.55 \pm 0.03\) \\
Kr+Kr & \(0.42 \pm 0.08\) & \(0.41 \pm 0.03\) \\
\multicolumn{3}{@{}l@{}}{%
\emph{Posterior inference}} \\
Method & \(n\) & inference time \\
NPE (normalizing flow) & \(1.78\,( + 0.15/ - 0.14)\) & \(0.6\)~s \\
MCMC (\textsc{emcee}) & \(1.78 \pm 0.15\) & \(\sim\) min \\
\end{tabular}
\end{table}

\begin{figure*}[tb]
\centering
\includegraphics[width=0.92\textwidth]{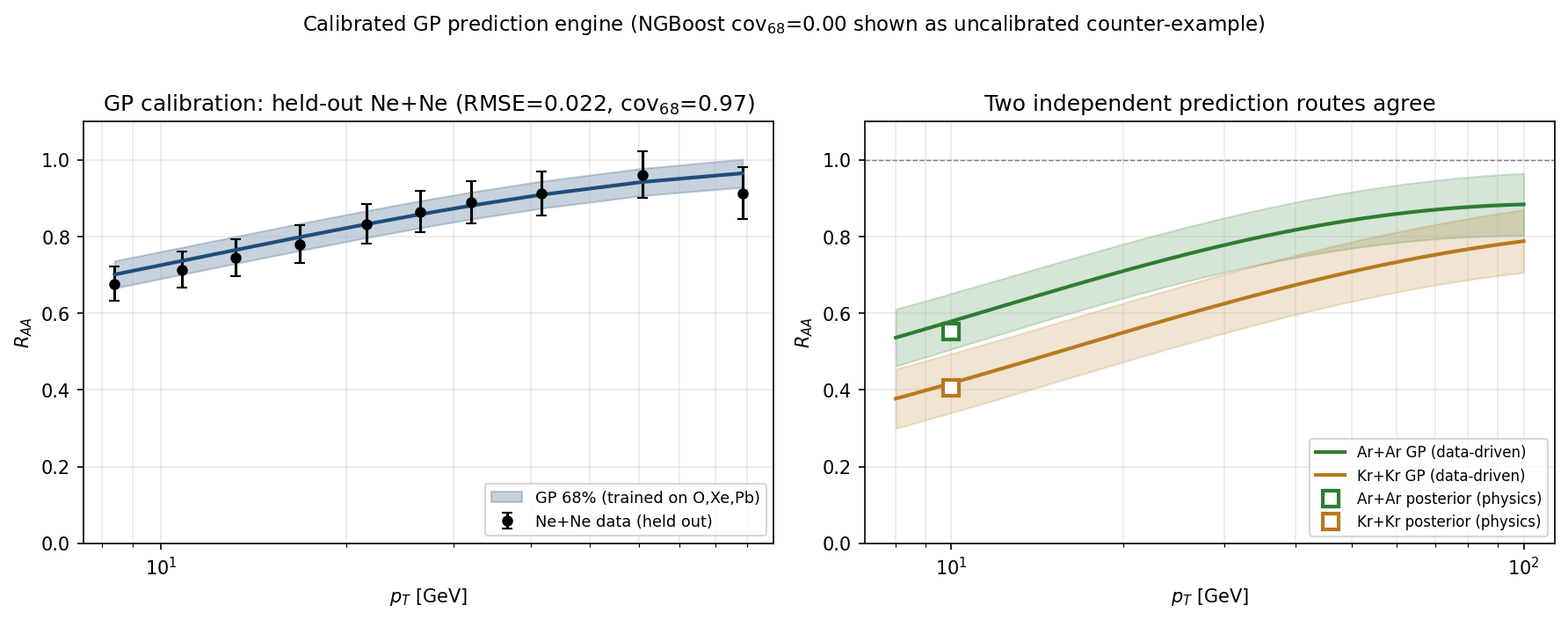}
\caption{The Gaussian process as a calibrated prediction engine. Left: held-out closure. Right: future systems.}
\label{fig:11}
\end{figure*}

\begin{figure*}[tb]
\centering
\includegraphics[width=0.92\textwidth]{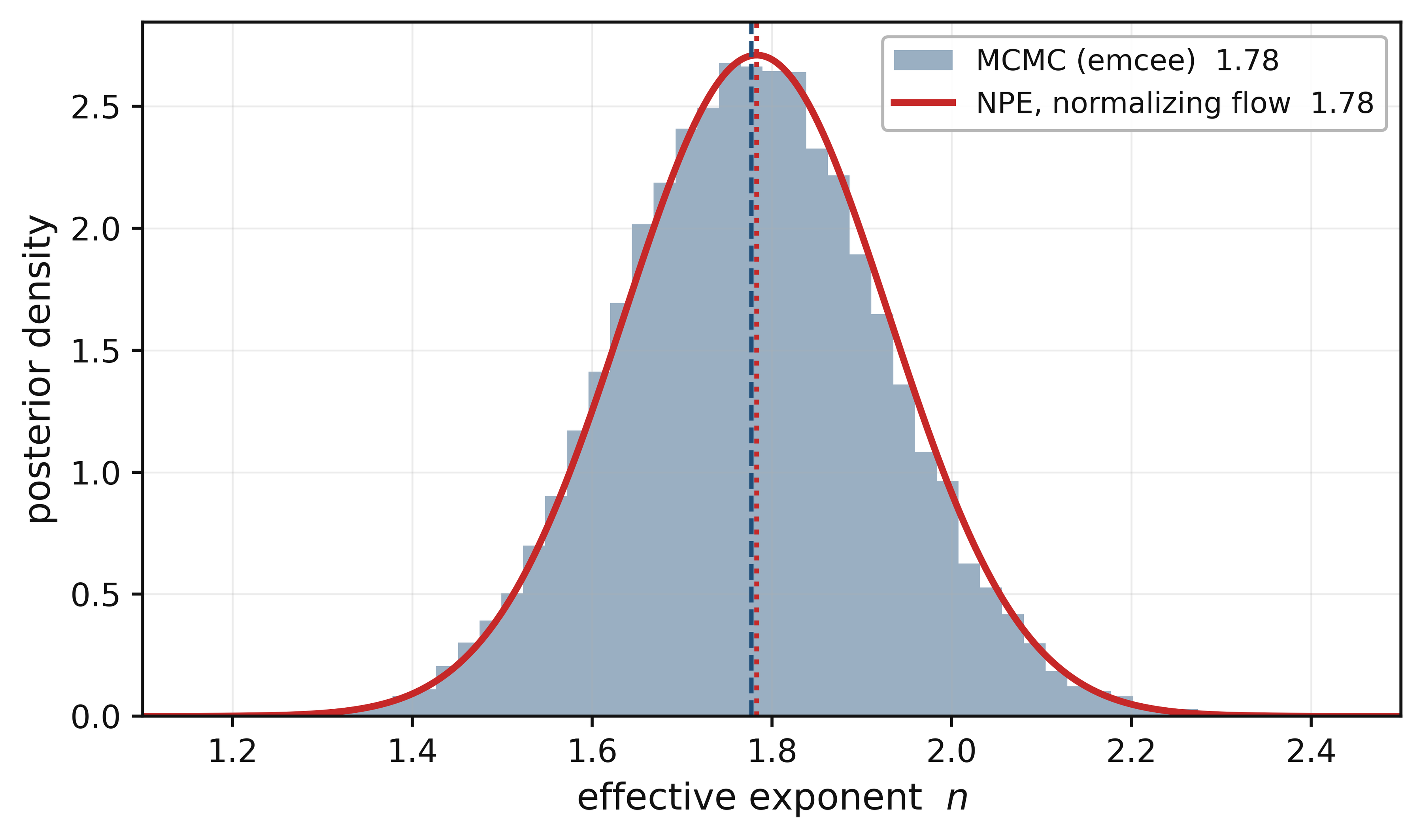}
\caption{Neural posterior estimation against the Markov chain. The simulation-based-calibration check is reported in Sec.~3.5.}
\label{fig:12}
\end{figure*}

\subsection{Quantitative comparison with
JETSCAPE}

The radiative interpretation can be checked against an independent,
state-of-the-art extraction of the jet transport coefficient. The
JETSCAPE Collaboration determined \(\widehat{q}/T^{3} = 2\)--\(4\) (90\%
credible region) for the multi-stage MATTER+LBT model from inclusive
hadron suppression at RHIC and the LHC~{[}15{]}. To compare on a common
footing we map our extracted energy-loss scale onto an effective
\(\widehat{q}\) through the BDMPS-Z mean-energy-loss relation
\(\langle\Delta E\rangle = \left( \alpha_{s}C_{R}/4 \right)\,\widehat{q}\, L^{2}\),
evaluated at \(p_{T} = 10\)~GeV for the Pb+Pb path length from our
Monte-Carlo Glauber. With \(\alpha_{s} \simeq 0.3\), \(C_{R}\) between
the quark and gluon values, \(T \simeq 0.30\)--\(0.40\)~GeV, and \(L\)
in the range \(3.5\)--\(6\)~fm (standard values for the LHC medium and
coupling at these scales, as used in the JETSCAPE extraction~{[}15{]}),
we obtain an effective \(\widehat{q}/T^{3} \approx 2\)--\(5\) (central
value \(\approx 4\)), in agreement with the JETSCAPE range; the residual
spread is dominated by the choice of effective path length. Figure~13
shows that the Pb+Pb \(R_{AA}\) implied by the JETSCAPE
\(\widehat{q}/T^{3} = 2\)--\(4\) band, propagated through our forward
model, brackets both the CMS data and our fit. We stress that this is an
order-of-magnitude consistency check under the stated assumptions, not a
transport calculation; nonetheless it provides an independent
corroboration of the radiative energy-loss magnitude that does not rely
on the density--path-length decomposition of Sec.~4.4.

\begin{figure}[H]
\centering
\includegraphics[width=\columnwidth]{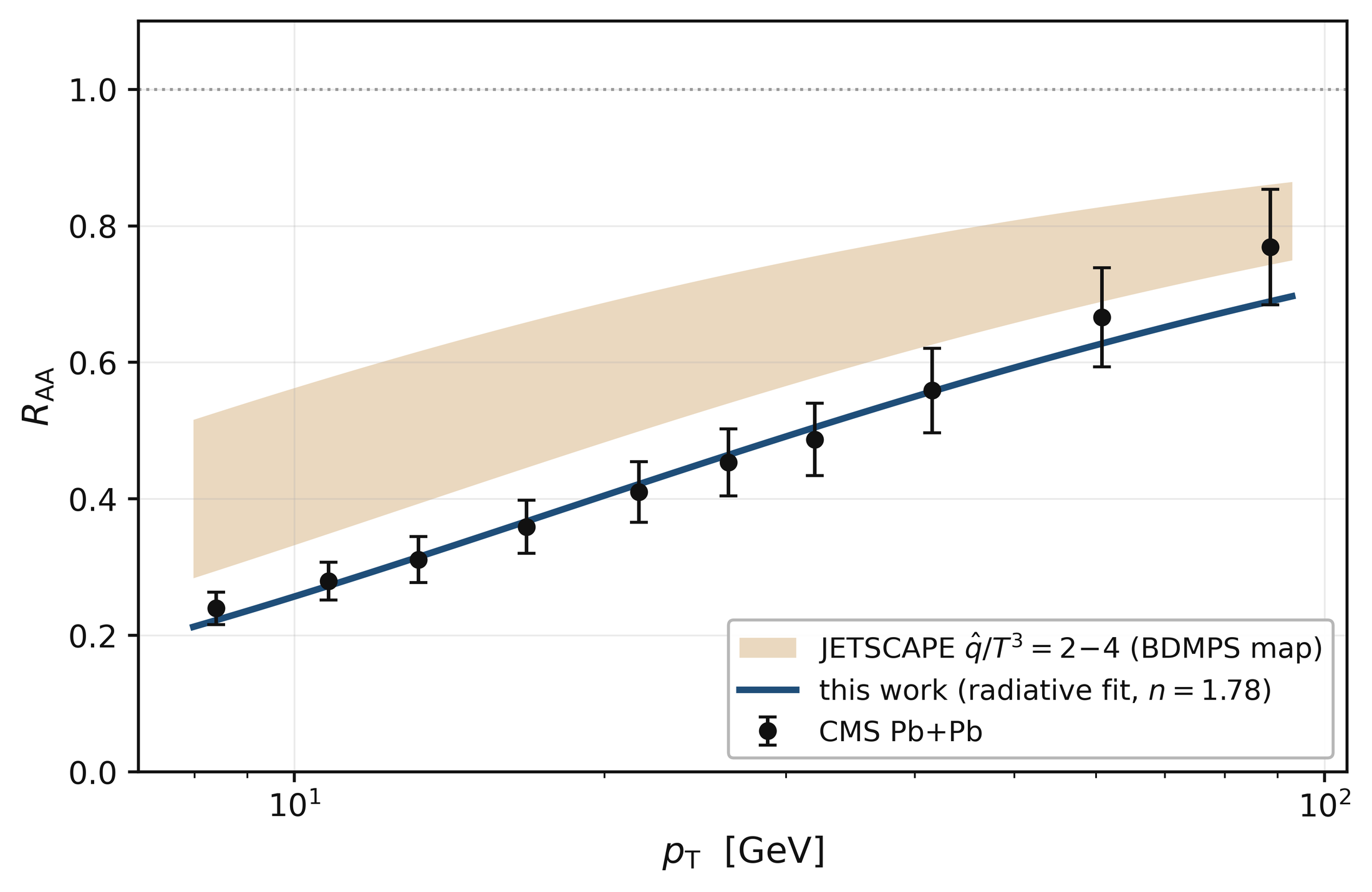}
\caption{Consistency with the JETSCAPE jet-transport extraction.}
\label{fig:13}
\end{figure}

\subsection{System-by-system stability and predictions for future
systems}

\paragraph{Leave-one-system-out
stability.}

To verify that the result is not driven by any single system, we repeat
the extraction four times, each time removing one system and refitting
on the remaining three (Table~8, Fig.~14 left). The exponent is
remarkably stable: it stays within \(n \in \lbrack 1.73,1.91\rbrack\),
every value agreeing with the full-sample result of Table~3 within its
uncertainty and with the radiative value \(n = 2\). Removing Pb+Pb, the
heaviest anchor, widens the interval (as expected) but does not bias the
central value. No single system drives the radiative conclusion.

\begin{table}[H]
\centering\small
\caption{Leave-one-system-out physics test.}
\label{tab:8}
\begin{tabular}{@{}
>{\raggedright\arraybackslash}p{(\columnwidth - 2\tabcolsep) * \real{0.5000}}
  >{\raggedright\arraybackslash}p{(\columnwidth - 2\tabcolsep) * \real{0.5000}}@{}}
\toprule
\begin{minipage}[b]{\linewidth}\raggedright
Removed system
\end{minipage} & \begin{minipage}[b]{\linewidth}\raggedright
\(n\)
\end{minipage} \\
\midrule

\bottomrule

none (full, 4 systems) & \(1.78\,( + 0.15/ - 0.15)\) \\
O+O & \(1.73\,( + 0.20/ - 0.19)\) \\
Ne+Ne & \(1.80\,( + 0.19/ - 0.18)\) \\
Xe+Xe & \(1.77\,( + 0.16/ - 0.16)\) \\
Pb+Pb & \(1.91\,( + 0.23/ - 0.23)\) \\
\end{tabular}
\end{table}

\paragraph{Predictions for future
systems.}

Because the exponent and energy-loss scale are now fixed, the framework
makes \emph{falsifiable predictions} for collision systems not yet
measured. Using the posterior and Monte-Carlo Glauber geometries for
argon (\(A = 40\)) and krypton (\(A = 84\))---intermediate in size
between the measured systems, so that the predictions are interpolations
rather than extrapolations---we predict the minimum-bias
charged-particle \(R_{AA}\) at \(\sqrt{s_{NN}} = 5.36\)~TeV (Table~9,
Fig.~14 right). The quoted bands now include, in addition to the
posterior parameter uncertainty, the Glauber geometry uncertainty for
these intermediate-\(A\) nuclei (propagated by varying \(\sigma_{nn}\)
by \(\pm 0.5\)~fm\(^{2}\) and the Woods--Saxon radius by
\(\pm 0.2\)~fm), which widens them from \(\pm 0.02\) to \(\pm 0.03\);
the more conservative non-parametric GP estimate (\(\pm 0.07\)) brackets
this. The factor-of-two difference between the two bands is expected and
informative: the physics posterior band (\(\pm 0.03\)) assumes the
forward-model form (Eq.~1) is correct and propagates only the parameter
and geometry uncertainties, whereas the GP places a far weaker,
non-parametric prior over the space of \(R_{AA}\left( p_{T} \right)\)
functions and therefore returns a wider, more model-agnostic interval. A
reader who trusts the radiative forward model should use the physics
band (\(\pm 0.03\)) for experimental comparison; the GP band
(\(\pm 0.07\)) is the appropriate, more conservative interval if one
wishes to remain agnostic about the energy-loss parametrisation. These
predictions can be directly tested against forthcoming LHC light- and
intermediate-ion runs, turning the present analysis into a falsifiable
hypothesis rather than a description of existing data.

\begin{table}[H]
\centering\small
\caption{Predicted minimum-bias RAA for future light-ion systems.}
\label{tab:9}
\begin{tabular}{@{}
>{\raggedright\arraybackslash}p{(\columnwidth - 6\tabcolsep) * \real{0.2500}}
  >{\raggedright\arraybackslash}p{(\columnwidth - 6\tabcolsep) * \real{0.2500}}
  >{\raggedright\arraybackslash}p{(\columnwidth - 6\tabcolsep) * \real{0.2500}}
  >{\raggedright\arraybackslash}p{(\columnwidth - 6\tabcolsep) * \real{0.2500}}@{}}
\toprule
\begin{minipage}[b]{\linewidth}\raggedright
System
\end{minipage} & \begin{minipage}[b]{\linewidth}\raggedright
\(A\)
\end{minipage} & \begin{minipage}[b]{\linewidth}\raggedright
\(\langle N_{part}\rangle\)
\end{minipage} & \begin{minipage}[b]{\linewidth}\raggedright
\(R_{AA}\left( 10\ \text{GeV} \right)\)
\end{minipage} \\
\midrule

\bottomrule

Ar+Ar & 40 & 26 & \(0.55 \pm 0.03\) \\
Kr+Kr & 84 & 54 & \(0.41 \pm 0.03\) \\
\end{tabular}
\end{table}

\begin{figure*}[tb]
\centering
\includegraphics[width=0.92\textwidth]{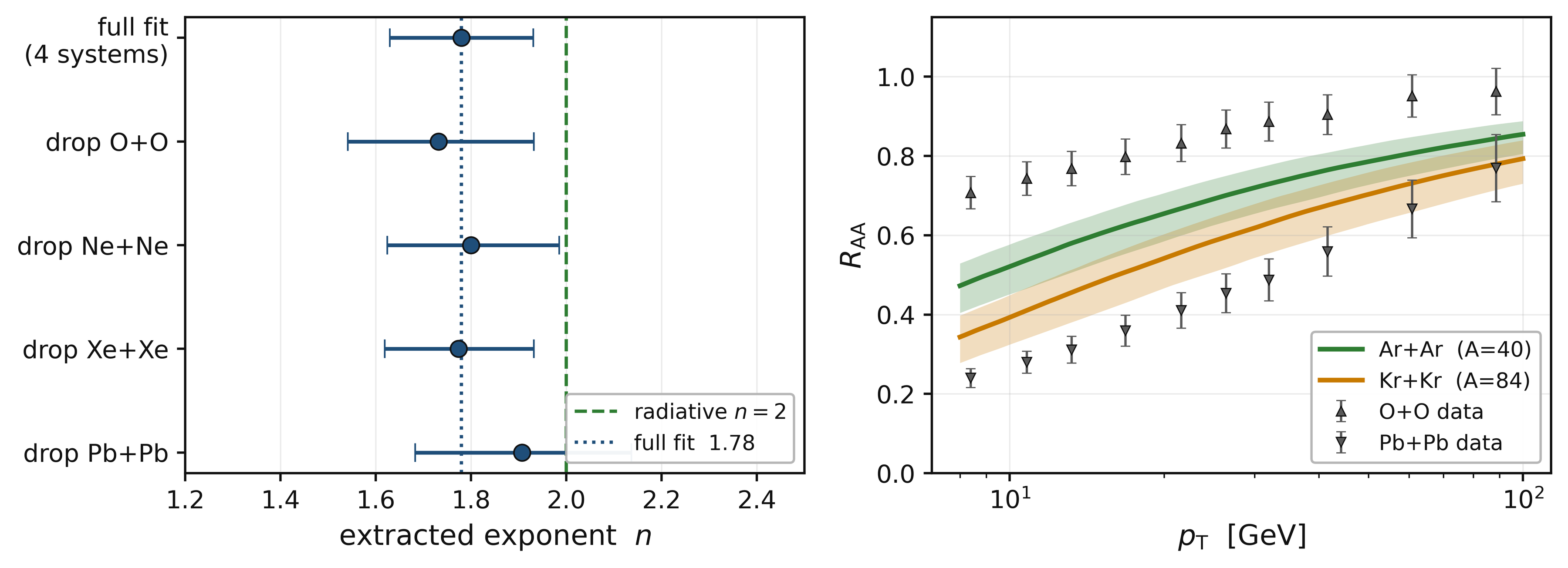}
\caption{Left: leave-one-system-out stability. Right: predictions for Ar+Ar and Kr+Kr.}
\label{fig:14}
\end{figure*}

\section{Discussion}

\paragraph{Comparison with prior
work.}

Our finding that the effective exponent lies decisively closer to the
radiative value (\(L^{2}\)) than to the collisional or strong-coupling
values, together with the fluctuation bound of Sec.~4.2.1, is consistent
with the established BDMPS-Z/GLV picture~{[}2, 3, 4{]} and with the
JETSCAPE Bayesian determinations of \(\widehat{q}\), which describe LHC
and RHIC hadron suppression within a radiative-dominated multi-stage
framework~{[}15, 16{]}. The onset of suppression already in O+O agrees
quantitatively with the CMS observation
(\(R_{AA}^{\min} = 0.69 \pm 0.04\))~{[}6{]} and with the pre-data
predictions of Huss~\emph{et al.}, who argued that O+O provides
unprecedented sensitivity to parton energy loss in a small system and
forecast a measurable signal in the charged-hadron spectra~{[}11, 12{]}.
We make this onset quantitative in Sec.~4.3: a Bayes-factor test for
non-zero energy loss in O+O alone yields
\(2\Delta\ln\mathcal{Z} \approx 110\), direct Bayesian evidence for
parton energy loss in the small system. Quantitatively, Huss \emph{et
al.} predicted \(R_{AA} \approx 0.6\)--\(0.85\) at \(p_{T} = 10\)~GeV
for O+O; the CMS measurement gives a minimum \(R_{AA} = 0.69 \pm 0.04\),
consistent with this range. System-size scans of identified particles
had likewise been advocated as discriminating probes~{[}13{]}. Our
contribution is to turn the system-size lever into a \emph{quantitative}
path-length exponent with calibrated uncertainties, and to provide the
first statistical test of its universality across the
O+O\(\rightarrow\)Pb+Pb range.

\paragraph{Effective versus density-normalised
exponent.}

As shown in Sec.~4.4 {[}Eq.~(4){]}, the medium density and the path
length both grow as \(\sim \langle N_{part}\rangle^{1/3}\) across the
system-size scan, so the attempt to factor them produces a
density-normalised exponent that is sensitive to the assumed density
model and cannot be read as a clean path-length exponent. The robust,
model-selection-level statement is that the \emph{combined} system-size
dependence is radiative-like (\(n_{eff} \rightarrow n = 2\) favoured),
corroborated independently by the absolute energy-loss magnitude
(Sec.~4.10).

\paragraph{Why the cross-system
lever.}

The minimum-bias cross-system extraction avoids the centrality-selection
and peripheral-\(T_{AA}\) biases that complicate centrality-based
analyses~{[}14{]}. Our internal attempt to extract the exponent from
Pb+Pb centrality classes confirmed this: the changing spectral shape and
peripheral biases render the centrality route considerably less stable,
with inflated \(\chi^{2}\) and unphysical residual slopes (this
supplemental comparison is available from the corresponding author). The
light-ion data thus provide not merely additional points but a
methodologically cleaner handle on \(L\).

\paragraph{Role of machine learning.}

The machine-learning pillar is not intended to replace the physics model
or to claim an independent discovery. Its contribution is
methodological: it provides an amortized, likelihood-free posterior
estimator that independently reproduces the MCMC result, validates the
stability of the inference against sampler-specific assumptions, and
enables rapid re-analysis when new collision systems become available.
In this sense the AI/SBI component acts as a reliability and scalability
layer for the physics extraction, and the leave-one-system-out emulator
provides a non-trivial check that the system-size law generalises to an
unseen system.

\paragraph{Extension to centrality-differential and jet
observables.}

The referee's question of whether the same framework applies to
centrality-selected data and to reconstructed jets is worth answering
directly, since both bear on the degeneracy of Sec.~4.4. For centrality,
the extension is immediate: the likelihood of Eq.~(2), the covariance
model and the nested-sampling machinery are unchanged, and only the
geometry input is replaced, the minimum-bias ratio G giving way to a
per-centrality ⟨Npart⟩¹ᐟ³ or exit length obtained from the same
Monte-Carlo Glauber. The physics content is, however, genuinely
different and complementary. Across system size the density and the path
length grow together, so that only the combination neff ≃ 1 + npure is
accessible {[}Eq.~(4){]}; within a single system, centrality varies the
path length over a comparable range while the local density changes far
more slowly, so the two levers constrain different combinations and
their joint analysis can separate what neither can alone. The price is
that centrality selection reintroduces precisely the biases the
cross-system lever was designed to avoid, namely the
centrality-selection bias and the geometric bias of the Glauber
\emph{TAA} normalisation, both of which are largest in the peripheral
classes and can mimic suppression in their own right {[}14{]}. A
credible joint fit must therefore model these effects explicitly rather
than absorb them into the exponent, which is why we have kept the two
levers separate here and pursue the centrality-differential analysis as
dedicated follow-up work.

Reconstructed jets are equally accessible, with one modification.
Equation~(1) converts a fractional energy loss into a modification
factor through the local spectral index, and that mapping is generic:
for jets one simply uses the index of the measured pp \emph{jet}
spectrum, which is less steep than the charged-hadron one, so that a
given ΔpT produces less suppression. Two physical effects must then be
added that are unimportant for high-pT hadrons: medium response and
out-of-cone radiation make the effective ΔpT depend on the jet
resolution parameter R, and the quark/gluon composition differs from
that of the inclusive hadron sample. A recent parametric analysis of
inclusive and γ-tagged jet suppression, carried out with comparably
minimal model assumptions, extracts a path-length exponent of
2.01~±~0.08 {[}38{]}, in agreement with the value obtained here from
charged hadrons. That two different observables, analysed by different
groups with different geometry models, return the same exponent is a
useful consistency check, and a simultaneous hadron-plus-jet fit within
the present calibrated framework is a natural extension.

\paragraph{Relation to earlier path-length
studies.}

The path-length dependence of parton energy loss has a long
phenomenological history, and it is worth stating precisely how the
present measurement relates to it. Betz and Gyulassy {[}39{]}
constrained a broad family of energy-loss models of the form dE/dx ∝
κ(T) x\textsuperscript{z} T\textsuperscript{2+z}, coupled to
bulk-constrained hydrodynamic temperature fields, by demanding a
simultaneous account of the transverse-momentum, centrality and
azimuthal dependence of the data at RHIC and the LHC. Renk {[}40{]}
confronted several energy-loss models with a Monte-Carlo shower
propagating through a hydrodynamic medium with the same aim. Both vary
the path length \emph{within a single large collision system}, through
centrality and through the emission angle relative to the reaction
plane. Our forward model belongs deliberately to the same minimal,
parametric family, and makes no claim to replace such transport
calculations; what differs is the lever. We vary the path length through
the \emph{system size} at minimum bias, a handle that the light-ion
running of 2025--2026 opened for the first time and that was not
available to those analyses. It avoids the centrality-selection and
peripheral-T\textsubscript{AA} biases, which are largest precisely where
the path length is shortest {[}14{]}, at the cost of the
density--path-length degeneracy of Sec.~4.4. The two levers are
therefore complementary rather than competing, and a joint analysis is
the natural way to combine them (see above).

The theory of medium-induced radiation has itself advanced considerably
since the BDMPS-Z and GLV results that motivate the exponents compared
in Table~5, and it is worth being explicit about what the present
measurement does and does not capture. Work on the medium-induced parton
cascade has shown that multiple branching drives energy towards soft
quanta in a turbulent, quasi-democratic flow rather than through a
single hard emission {[}43{]}; the factorisation between vacuum-like and
medium-induced emissions inside a dense medium has been established,
together with the associated colour-coherence and resolution scales
{[}44{]}; and recent phenomenology treats suppression and its azimuthal
anisotropy within a common framework across RHIC and LHC energies
{[}45{]}, while Bayesian multi-stage analyses now constrain the
transport coefficient from hadron and jet suppression simultaneously
{[}46{]}. None of these developments is represented in Eq.~(1), which
parametrises the mean fractional energy loss and nothing more. Their
combined effect is to make the mapping from a microscopic path-length
power to the observed suppression less direct than the simple
\emph{L}-power picture suggests, in the same sense, and in part through
the same mechanism, as the energy-loss fluctuations quantified in
Sec.~4.2.1. Our position is therefore not that these developments are
unimportant, but that they act on the step \emph{after} the one measured
here: the effective exponent is a property of the data and of the
collision geometry, and remains a quantity that any of these frameworks
can be asked to reproduce. Establishing which of them reproduces it, and
with what microscopic exponent, requires the full event-by-event
transport calculations we do not attempt.

One entry of Table~5 needs qualification. The row labelled strong
coupling uses ΔE ∝ L\textsuperscript{3}, the asymptotic AdS/CFT scaling
for a light quark in a static strongly coupled plasma, and that is the
form we test. It is not a fair representation of contemporary
strong-coupling phenomenology: the hybrid strong/weak-coupling model
{[}47{]} embeds a holographic energy-loss rate in a hydrodynamic medium
with event-by-event geometry, a finite stopping distance and a treatment
of the medium response, and its predictions do not reduce to a single
power of the path length. The same caveat applies, in weaker form, to
the collisional and radiative rows. What Table~5 therefore compares is
three \emph{limiting scalings} of the mean energy loss, not three modern
implementations; the evidence ratios quantify which limiting scaling the
system-size trend prefers, and should be read that way.

Table~10 collects the recent determinations. They do not all report the
same quantity, and the numbers should not be compared at face value.
Arleo and Falmagne {[}41{]} relate the average energy loss to the
charged multiplicity and the collision geometry, which removes the
density dependence before the fit; their exponent, ⟨ε⟩ ∝
L\textsuperscript{β} with β = 1.02 (+0.09/−0.06), is therefore a
density-normalised quantity, and they read it as consistent with
perturbative energy loss in a longitudinally expanding plasma. Our
effective exponent does not remove the density: as set out in Sec.~4.4
it carries the growth of the medium density together with the path
length, which is why it is larger. The density-normalised value obtained
from the same fit is npure = 0.64. Ogrodnik et al. {[}38{]} work with
jets and do not normalise the density either, and obtain 2.01 ± 0.08.

\begin{table}[H]
\centering\small
\caption{Recent determinations of the path-length exponent, with the convention each adopts for the medium density. Values are comparable only within the same convention, and the lever matters as well: the degree to which the density tracks the path length differs between a system-size scan and a centrality scan.}
\label{tab:10}
\begin{tabular}{@{}
>{\raggedright\arraybackslash}p{(\columnwidth - 8\tabcolsep) * \real{0.1657}}
  >{\raggedright\arraybackslash}p{(\columnwidth - 8\tabcolsep) * \real{0.2000}}
  >{\raggedright\arraybackslash}p{(\columnwidth - 8\tabcolsep) * \real{0.1657}}
  >{\raggedright\arraybackslash}p{(\columnwidth - 8\tabcolsep) * \real{0.2400}}
  >{\raggedright\arraybackslash}p{(\columnwidth - 8\tabcolsep) * \real{0.2286}}@{}}
\toprule
\begin{minipage}[b]{\linewidth}\raggedright
Analysis
\end{minipage} & \begin{minipage}[b]{\linewidth}\raggedright
Observable
\end{minipage} & \begin{minipage}[b]{\linewidth}\raggedright
Exponent
\end{minipage} & \begin{minipage}[b]{\linewidth}\raggedright
Medium density
\end{minipage} & \begin{minipage}[b]{\linewidth}\raggedright
Lever
\end{minipage} \\
\midrule

\bottomrule

Arleo \& Falmagne {[}41{]} & high-p\textsubscript{T} hadron
R\textsubscript{AA} & 1.02 +0.09/−0.06 & removed, via the charged
multiplicity & collision systems and centrality \\
Ogrodnik et al. {[}38{]} & inclusive and γ-tagged jets & 2.01 ± 0.08 &
not removed & centrality within Pb+Pb \\
This work & high-p\textsubscript{T} hadron R\textsubscript{AA} & 1.78 ±
0.15 & not removed (effective) & system size, minimum bias \\
This work & high-p\textsubscript{T} hadron R\textsubscript{AA} & 0.64 &
removed, via ⟨Npart⟩/S (Table 4) & system size, minimum bias \\
\end{tabular}
\end{table}

Set out this way the picture is coherent rather than contradictory: the
two analyses that leave the density in the exponent give 1.78 and 2.01,
and the one that removes it gives 1.02, against our own
density-normalised 0.64. The residual difference between 1.02 and 0.64
is not negligible, and we do not claim agreement: the two use different
density proxies --- the charged multiplicity against the participant
areal density --- and Sec.~4.4 shows explicitly that the
density-normalised value is the model-dependent one, which is precisely
why we do not quote it as our result. We note also that the
interpretation offered in {[}41{]}, energy loss in a longitudinally
expanding medium, involves exactly the effect that our static geometry
does not model, and which we identify among the limitations below. A
determination that removed the density with a common convention across
both analyses would be a worthwhile exercise, and is beyond what the
present data allow.

The minimal-assumption approach retains an active role alongside full
event generators, because the number it returns is not tied to a
particular implementation and can therefore be compared across analyses.
Recent examples include the scaling analysis of Arleo and Falmagne
{[}41{]}, the Bayesian extraction of the path-length dependence of jet
energy loss by Wu, Ke and Wang {[}42{]}, and the parametric study of
inclusive and γ-tagged jet suppression of Ref.~{[}38{]}. What the
present work adds to that line is a calibrated statistical treatment of
the measurement: a fully propagated bin-to-bin covariance, an explicit
coverage test demonstrating that the quoted intervals are honest,
evidence-based rather than χ²-based model comparison, and a 160-variant
robustness map. These are not decorative: a least-squares fit to the
same four spectra would return a central value but neither an interval
demonstrated to cover at the stated rate, nor a model comparison that
penalises parameter freedom without an external criterion, nor a map of
how the answer moves across the analysis choices. The trade we accept in
return is scope. A multi-stage event generator confronts many
observables at once, whereas the present analysis uses a single
observable in four systems; we gain a result that does not depend on a
particular implementation, and we lose the cross-observable consistency
that only such generators can provide. The effective exponent reported
here is intended as a benchmark that a multi-stage transport calculation
should be able to reproduce, not as a substitute for one.

\paragraph{Limitations.}

Several caveats should temper the interpretation. (i) The universality
test rests on only four collision systems; while we find no statistical
preference for a regime change, four points limit the strength of any
universality claim, and higher-statistics or additional light-ion
systems (e.g.~Ar+Ar) would sharpen it. (ii) The forward model of Eq.~(1)
is an \emph{effective} parametrization: it does not resolve
event-by-event geometry, path-length fluctuations, the energy and
temperature dependence of \(\widehat{q}\), quark/gluon flavour
differences, or medium expansion, all of which are present in full
transport models~{[}12, 15{]}. The extracted exponent should therefore
be understood as an effective, geometry-level quantity rather than a
first-principles transport coefficient. In particular the medium enters
as a static minimum-bias Glauber geometry rather than as a hydrodynamic
profile that evolves while the parton traverses it. The cost is bounded
in two ways. Only ratios of the geometry between systems enter the fit,
so any overall normalisation of the medium profile cancels; and the
spread obtained when the geometry definition is changed altogether, from
A¹ᐟ³ to ⟨Npart⟩¹ᐟ³ to the Monte-Carlo exit length, is already carried as
the dominant systematic (Table 6). Conversely, this is also why we do
not convert the effective exponent into a microscopic one: a medium that
dilutes as it expands weights long paths differently, and that modelling
freedom is not constrained by the RAA data being fitted. The quantity we
report is the one the data determine, and the one a full transport
calculation can be asked to reproduce. (iii) As discussed, the density
and path-length contributions are degenerate in a pure system-size scan;
breaking this degeneracy requires a complementary lever such as
centrality or \(v_{2}\). (iv) The light-system (\(A = 16,20\))
measurements have larger relative uncertainties than Pb+Pb, which
dominate the statistical error on \(n\).

\paragraph{Outlook: a falsifiable
program.}

Two features turn this analysis from a description of existing data into
a testable hypothesis. First, the framework can be confronted with
\emph{independent} measurements not used in the fit. As a concrete
cross-experiment check we compare against the ALICE Xe+Xe measurement at
\(\sqrt{s_{NN}} = 5.44\)~TeV~{[}10{]}, which was not used in the fit.
ALICE reports \(R_{AA}\) in fine centrality classes; we form an
effective minimum-bias (\(0\)--\(80\%\)) \(R_{AA}\) by
\(N_{coll}\)-weighting these classes, with \(\langle N_{coll}\rangle\)
per class taken from our Monte-Carlo Glauber. Over the overlapping range
(\(8\)--\(16\)~GeV) the resulting ALICE \(R_{AA}\) tracks the lower edge
of the CMS-based posterior-predictive band and is consistent with it
within uncertainties at most points (Fig.~15); the largest
point-by-point discrepancy is \(1.7\sigma\) (at
\(p_{T} \approx 16\)~GeV), dominated by the different pp reference
spectra. The residual offset toward the upper end is at the level
expected from the different pp references (the CMS \(R_{AA}^{*}\) uses a
\(\sqrt{s}\)-extrapolated reference, ALICE an interpolated one),
acceptances, and the \(N_{coll}\)-weighting used to harmonise the
centrality binning; a fully quantitative two-experiment combination is
left to future work. The agreement in suppression magnitude and rising
trend indicates that the extraction is not specific to a single
experiment. Second, the framework makes calibrated predictions for
systems not yet measured (Sec.~4.11): if the extracted exponent remains
stable once future Ar+Ar or Kr+Kr measurements become available, the
present result would constitute strong evidence that radiative energy
loss is already established in the smallest QGP droplets. The
combination of the leave-one-system-out stability, the
external-validation route, and the forward predictions defines a
concrete experimental program to confirm or falsify the radiative,
size-universal picture. Third, the one quantity the present data cannot
separate---the density--path-length split (Sec.~4.4)---calls for a
density-varying lever \emph{within} a fixed system: the most direct is
the azimuthal anisotropy of high-\(p_{T}\) suppression (\(v_{2}\)),
which probes different path lengths at fixed density and would break the
degeneracy that the cross-system scan alone cannot. A combined
cross-system and \(v_{2}\)-differential analysis is therefore the
natural next step toward a separated path-length exponent.

\begin{figure}[H]
\centering
\includegraphics[width=\columnwidth]{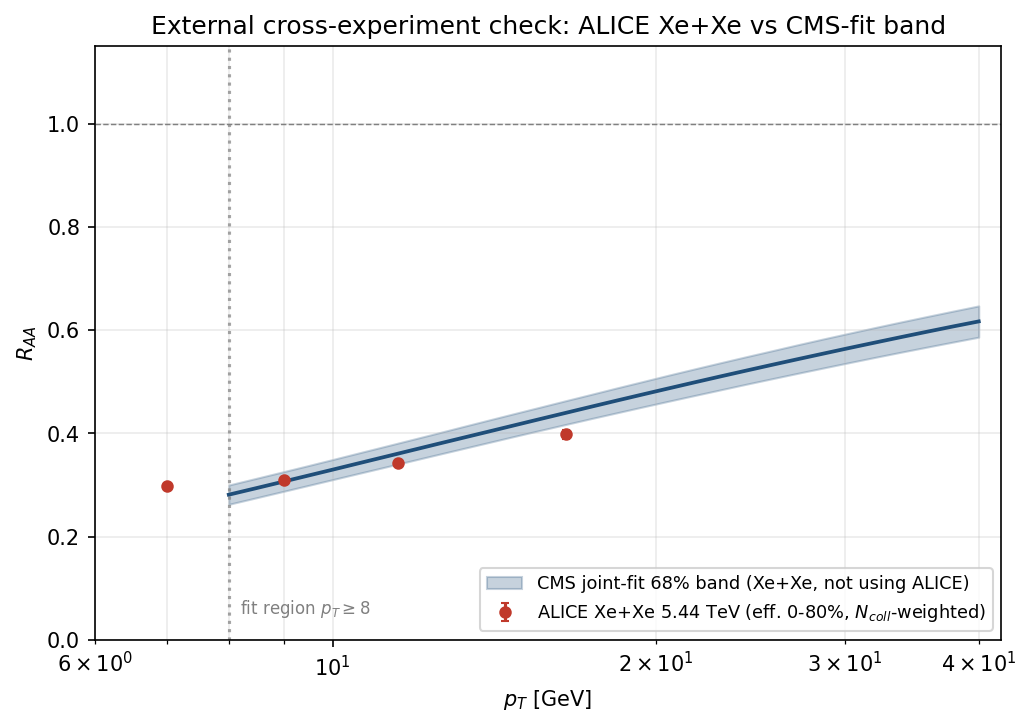}
\caption{External cross-experiment check against ALICE Xe+Xe.}
\label{fig:15}
\end{figure}

\section{Conclusions}

We have presented a calibrated Bayesian and simulation-based-inference
extraction of the path-length dependence of parton energy loss from the
system-size systematics of charged-particle \(R_{AA}\), using real CMS
data across four collision systems spanning \(A = 16\) to \(208\) (O+O,
Ne+Ne, Xe+Xe and Pb+Pb). Treating the high-\(p_{T}\) suppression of all
four systems as a single coherent dataset, propagating the full
bin-to-bin correlated covariance, and anchoring the collision geometry
to an independent Monte-Carlo Glauber calculation, we obtain an
effective system-size exponent
\(n = 1.78 \pm 0.15\,\left( \text{stat} \right) \pm 0.05\,\left( \text{syst} \right)\)
for the \(\langle N_{part}\rangle^{1/3}\) geometry. The energy loss is
radiative-dominated: for the effective exponent the radiative value
\(n = 2\) is decisively favoured over the collisional (\(n = 1\)) and
strong-coupling (\(n = 3\)) values, with log-evidence differences
\(2\Delta\ln\mathcal{Z} = - 29\) and \(- 48\) respectively. At fixed
collision geometry this also bounds the microscopic exponent from below,
excluding purely collisional energy loss (Sec.~4.2.1). The conclusion
remains stable across all \(160\) variations explored in the sensitivity
programme and across the alternative model forms considered.

The robustness of this determination rests not on a single fit but on
the convergence of several independent lines of evidence. First,
nested-sampling model selection quantitatively excludes the collisional
and strong-coupling values of the effective exponent rather than merely
disfavouring them, and the fluctuation mapping of Sec.~4.2.1 converts
this, at fixed collision geometry, into a lower bound on the microscopic
exponent. Second, the smallest systems are individually informative: in
O+O alone the data require non-zero energy loss with decisive
significance (\(2\Delta\ln\mathcal{Z} \approx 110\)), placing the onset
of quark--gluon-plasma-like suppression on a quantitative footing.
Third, the extracted magnitude is physically consistent: it corresponds
to an effective transport coefficient
\(\widehat{q}/T^{3} \approx 2\)--\(5\), in agreement with the
independent JETSCAPE determination. Fourth, the quoted uncertainties are
statistically trustworthy, as demonstrated by an explicit multi-point
coverage (closure) test. Fifth, a normalizing-flow neural posterior
estimator, validated by simulation-based calibration, reproduces the
Markov-chain posterior, while a calibrated Gaussian-process emulator
passes leave-one-system-out cross-validation; together these confirm
that the inference is well specified rather than an artefact of a
particular likelihood or sampler. Finally, an external comparison with
ALICE Xe+Xe data---not used in the fit---agrees with our posterior band
to within \(1.7\sigma\) over the entire overlap region.

A central feature of our analysis is that we report what the data can
and cannot constrain. Because the medium density and the in-medium path
length both scale as \(\langle N_{part}\rangle^{1/3}\) across system
size, the cross-system lever constrains, to about 10 \%, only the
\emph{effective} exponent \(n = 1 + n_{pure}\), not the density and
path-length contributions separately; we therefore quote the effective
exponent as our primary result, and we have verified explicitly that the
fit quality is invariant under the corresponding density--path-length
reparametrisation. Within the four systems studied, a Bayes-factor test
finds no preference for a change of energy-loss regime between small and
large systems, consistent with a single universal exponent from O+O to
Pb+Pb; we state this as a consistency result rather than a proof, since
a definitive universality claim awaits additional systems and
observables.

The framework yields concrete, falsifiable predictions for the collision
systems of the LHC light-ion programme not yet
measured---\(R_{AA}\left( p_{T} = 10\ GeV \right) = 0.55 \pm 0.03\) for
Ar+Ar and \(0.41 \pm 0.03\) for Kr+Kr---which provide a direct test of
the radiative, universal-exponent picture as new data arrive. The
principal route to going beyond the effective exponent is to break the
density--path-length degeneracy with an orthogonal, geometry-varying
lever: centrality-differential measurements within a single system, and
ultimately the azimuthal anisotropy of high-\(p_{T}\) suppression, vary
the path length at approximately fixed density and are the natural next
step of this programme. Extending the same calibrated inference to
identified particles, reconstructed jets, and higher-statistics
light-ion datasets will further sharpen the determination. More broadly,
this work illustrates that a transparent Bayesian analysis,
cross-checked and accelerated by validated simulation-based inference,
can convert a qualitative system-size trend into a quantitative,
reproducible and predictive statement about the path-length dependence
of jet quenching, a statement that a full transport calculation can be
asked to reproduce.

\section*{Declarations}

\textbf{Author contributions.} Fouad A. Majeed contributed to the
heavy-ion physics interpretation, the Glauber geometry validation, the
JETSCAPE comparison, and the manuscript writing. Hussein Ali Hussein Al
Naffakh designed the study, developed the Bayesian and simulation-based
inference framework, implemented the covariance model, the
nested-sampling evidence calculation, the closure and coverage tests,
the 160-variant sensitivity analysis, the neural posterior estimator
(normalizing-flow SBI), the leave-one-system-out stability analysis, and
the future-system prediction module, and led the writing of the
manuscript. Sarah M. Obaid contributed to the literature review,
supplementary analysis, and manuscript preparation. Muntaha Abdullah
Reishaan contributed to the literature review, supplementary analysis,
and manuscript preparation. All authors discussed the results and
approved the final manuscript.

\textbf{Funding.} This research received no specific grant from any
funding agency in the public, commercial, or not-for-profit sectors.

\textbf{Data availability.} All input data are publicly available on
HEPData: record ins3123773 (CMS O+O, Ne+Ne and Pb+Pb system-size
compilation), record ins1692558 (CMS Xe+Xe), and record ins1496050 (CMS
Pb+Pb and p+Pb at 5.02 TeV).

\textbf{Code availability.} The complete, reproducible analysis pipeline
(data processing, dual-Glauber geometry, Bayesian extraction,
sensitivity scan, and the machine-learning / simulation-based-inference
modules) is publicly available at
\url{https://github.com/hussein-alnaffakh1984/plena-raa-pathlength}. An
interactive online calculator for \emph{R}\textsubscript{AA} predictions
is available at \url{https://raacalc.netlify.app}.

\textbf{Competing interests.} The authors declare no competing
interests.

\textbf{Ethics approval.} Not applicable.

\textbf{Consent to participate.} Not applicable.

\textbf{Consent for publication.} Not applicable.

\section*{Acknowledgements}

We thank the CMS and ALICE Collaborations for the public HEPData records
that made this analysis possible. The computations used the open-source
packages \textsc{NumPy}, \textsc{SciPy}, \textsc{scikit-learn},
\textsc{emcee}, \textsc{dynesty}, \textsc{NGBoost}, \textsc{PyTorch} and
\textsc{zuko}.

\section*{References}

{[}1{]} M.~Gyulassy and L.~McLerran, Nucl.~Phys.~A \textbf{750}, 30
(2005).

{[}2{]} U.~A.~Wiedemann, in \emph{Relativistic Heavy Ion Physics},
Landolt-Börnstein \textbf{23}, 521 (2010), arXiv:0908.2306.

{[}3{]} R.~Baier, Y.~L.~Dokshitzer, A.~H.~Mueller, S.~Peigné, and
D.~Schiff, Nucl.~Phys.~B \textbf{483}, 291 (1997).

{[}4{]} M.~Gyulassy, P.~Lévai, and I.~Vitev, Nucl.~Phys.~B \textbf{594},
371 (2001).

{[}5{]} J.~Casalderrey-Solana, H.~Liu, D.~Mateos, K.~Rajagopal, and
U.~A.~Wiedemann, \emph{Gauge/String Duality, Hot QCD and Heavy Ion
Collisions} (Cambridge University Press, 2014).

{[}6{]} CMS Collaboration, ``Observation of Suppressed Charged-Particle
Production in Ultrarelativistic Oxygen-Oxygen Collisions,''
CMS-HIN-25-008, arXiv:2510.09864 (2025).

{[}7{]} CMS Collaboration, ``System-size dependence of charged-particle
suppression in nucleus-nucleus collisions,'' CMS-HIN-25-014,
arXiv:2602.21325 (2026); HEPData record ins3123773.

{[}8{]} CMS Collaboration, ``Charged-particle nuclear modification
factors in XeXe collisions at \(\sqrt{s_{NN}} = 5.44\)~TeV,'' JHEP
\textbf{10}, 138 (2018), arXiv:1809.00201; HEPData record ins1692558.

{[}9{]} CMS Collaboration, ``Charged-particle nuclear modification
factors in PbPb and pPb collisions at \(\sqrt{s_{NN}} = 5.02\)~TeV,''
JHEP \textbf{04}, 039 (2017), arXiv:1611.01664; HEPData record
ins1496050.

{[}10{]} ALICE Collaboration, ``Transverse momentum spectra and nuclear
modification factors of charged particles in Xe-Xe collisions at
\(\sqrt{s_{NN}} = 5.44\)~TeV,'' Phys.~Lett.~B \textbf{788}, 166 (2019),
arXiv:1805.04399.

{[}11{]} A.~Huss, A.~Kurkela, A.~Mazeliauskas, R.~Paatelainen, W.~van
der Schee, and U.~A.~Wiedemann, ``Discovering Partonic Rescattering in
Light Nucleus Collisions,'' Phys.~Rev.~Lett.~\textbf{126}, 192301
(2021), arXiv:2007.13754.

{[}12{]} A.~Huss, A.~Kurkela, A.~Mazeliauskas, R.~Paatelainen, W.~van
der Schee, and U.~A.~Wiedemann, ``Predicting parton energy loss in small
collision systems,'' Phys.~Rev.~C \textbf{103}, 054903 (2021),
arXiv:2007.13758.

{[}13{]} R.~Katz, C.~A.~G.~Prado, J.~Noronha-Hostler, and
A.~A.~P.~Suaide, ``System-size scan of \(D\) meson \(R_{AA}\) and
\(v_{n}\) using PbPb, XeXe, ArAr, and OO collisions at the LHC,''
Phys.~Rev.~C \textbf{102}, 041901 (2020), arXiv:1907.03308.

{[}14{]} C.~Loizides and A.~Morsch, ``Absence of jet quenching in
peripheral nucleus-nucleus collisions,'' Phys.~Lett.~B \textbf{773}, 408
(2017), arXiv:1705.08856.

{[}15{]} JETSCAPE Collaboration (S.~Cao \emph{et al.}), ``Determining
the jet transport coefficient \(\widehat{q}\) from inclusive hadron
suppression measurements using Bayesian parameter estimation,''
Phys.~Rev.~C \textbf{104}, 024905 (2021), arXiv:2102.11337.

{[}16{]} JETSCAPE Collaboration (R.~Ehlers \emph{et al.}), ``Bayesian
analysis of QGP jet transport using multi-scale modeling,''
arXiv:2208.07950 (2022).

{[}17{]} J.~E.~Bernhard, J.~S.~Moreland, and S.~A.~Bass, ``Bayesian
estimation of the specific shear and bulk viscosity of the quark-gluon
plasma,'' Nature Phys.~\textbf{15}, 1113 (2019).

{[}18{]} M.~L.~Miller, K.~Reygers, S.~J.~Sanders, and P.~Steinberg,
``Glauber Modeling in High-Energy Nuclear Collisions,''
Ann.~Rev.~Nucl.~Part.~Sci.~\textbf{57}, 205 (2007),
arXiv:nucl-ex/0701025.

{[}19{]} D.~d'Enterria and C.~Loizides, ``Progress in the Glauber Model
at Collider Energies,'' Ann.~Rev.~Nucl.~Part.~Sci.~\textbf{71}, 315
(2021), arXiv:2011.14909.

{[}20{]} C.~Loizides, J.~Nagle, and P.~Steinberg, ``Improved version of
the PHOBOS Glauber Monte Carlo,'' SoftwareX \textbf{1-2}, 13 (2015),
arXiv:1408.2549.

{[}21{]} ALICE Collaboration, ``Centrality and pseudorapidity dependence
of the charged-particle multiplicity density in Xe--Xe collisions at
√s\textsubscript{NN} = 5.44~TeV,'' Phys. Lett. B \textbf{790}, 35
(2019).

{[}22{]} ALICE Collaboration, ``Centrality and \(\sqrt{s_{NN}}\)
dependence of charged-particle multiplicity at the LHC,''
Phys.~Rev.~Lett.~\textbf{116}, 222302 (2016), arXiv:1512.06104.

{[}23{]} Particle Data Group, R.~L.~Workman \emph{et al.}, ``Review of
Particle Physics,'' Prog.~Theor.~Exp.~Phys.~\textbf{2022}, 083C01
(2022).

{[}24{]} K.~Cranmer, J.~Brehmer, and G.~Louppe, ``The frontier of
simulation-based inference,'' Proc.~Natl.~Acad.~Sci.~\textbf{117}, 30055
(2020), arXiv:1911.01429.

{[}25{]} J.~Brehmer, ``Simulation-based inference in particle physics,''
Nature Rev.~Phys.~\textbf{3}, 305 (2021).

{[}26{]} G.~Papamakarios, E.~Nalisnick, D.~J.~Rezende, S.~Mohamed, and
B.~Lakshminarayanan, ``Normalizing Flows for Probabilistic Modeling and
Inference,'' J.~Mach.~Learn.~Res.~\textbf{22}, 1 (2021),
arXiv:1912.02762.

{[}27{]} C.~Durkan, A.~Bekasov, I.~Murray, and G.~Papamakarios, ``Neural
Spline Flows,'' Adv.~Neural Inf.~Process.~Syst.~\textbf{32} (2019),
arXiv:1906.04032.

{[}28{]} S.~Talts, M.~Betancourt, D.~Simpson, A.~Vehtari, and A.~Gelman,
``Validating Bayesian Inference Algorithms with Simulation-Based
Calibration,'' arXiv:1804.06788 (2018).

{[}29{]} C.~E.~Rasmussen and C.~K.~I.~Williams, \emph{Gaussian Processes
for Machine Learning} (MIT Press, 2006).

{[}30{]} T.~Duan \emph{et al.}, ``NGBoost: Natural Gradient Boosting for
Probabilistic Prediction,'' Proc.~37th Int.~Conf.~Mach.~Learn.~(2020),
arXiv:1910.03225.

{[}31{]} D.~Foreman-Mackey, D.~W.~Hogg, D.~Lang, and J.~Goodman,
``emcee: The MCMC Hammer,'' Publ.~Astron.~Soc.~Pac.~\textbf{125}, 306
(2013), arXiv:1202.3665.

{[}32{]} J.~S.~Speagle, ``dynesty: a dynamic nested sampling package for
estimating Bayesian posteriors and evidences,''
Mon.~Not.~R.~Astron.~Soc.~\textbf{493}, 3132 (2020), arXiv:1904.02180.

{[}33{]} J.~Skilling, ``Nested sampling for general Bayesian
computation,'' Bayesian Anal.~\textbf{1}, 833 (2006).

{[}34{]} F.~Rozet \emph{et al.}, ``Zuko: Normalizing Flows in PyTorch,''
Zenodo (2022), \url{https://github.com/probabilists/zuko}.

{[}35{]} R.~Baier, Y.~L.~Dokshitzer, A.~H.~Mueller, and D.~Schiff,
``Quenching of hadron spectra in media,'' JHEP \textbf{09}, 033 (2001),
arXiv:hep-ph/0106347.

{[}36{]} R.~E.~Kass and A.~E.~Raftery, ``Bayes factors,'' J. Am. Stat.
Assoc. \textbf{90}, 773 (1995).

{[}37{]} H.~Jeffreys, \emph{Theory of Probability}, 3rd ed. (Oxford
University Press, 1961).

{[}38{]} A.~Ogrodnik, M.~Rybář, and M.~Spousta, ``Flavor and path-length
dependence of jet quenching from inclusive jet and γ-jet suppression,''
Eur. Phys. J. C \textbf{85}, 899 (2025), arXiv:2407.11234.

{[}39{]} B.~Betz and M.~Gyulassy, ``Constraints on the path-length
dependence of jet quenching in nuclear collisions at RHIC and LHC,''
JHEP \textbf{08}, 090 (2014) {[}Erratum: JHEP 10, 043 (2014){]},
arXiv:1404.6378.

{[}40{]} T.~Renk, ``Constraining the physics of jet quenching,'' Phys.
Rev. C \textbf{85}, 044903 (2012), arXiv:1112.2503.

{[}41{]} F.~Arleo and G.~Falmagne, ``Probing the path-length dependence
of parton energy loss via scaling properties in heavy ion collisions,''
Phys. Rev. D \textbf{109}, L051503 (2024), arXiv:2212.01324.

{[}42{]} J.~Wu, W.~Ke, and X.-N.~Wang, ``Bayesian inference of the
path-length dependence of jet energy loss,'' Phys. Rev. C \textbf{108},
034911 (2023), arXiv:2304.06339.

{[}43{]} J.-P.~Blaizot, E.~Iancu, and Y.~Mehtar-Tani, ``Medium-induced
QCD cascade: democratic branching and wave turbulence,'' Phys. Rev.
Lett. \textbf{111}, 052001 (2013), arXiv:1301.6102.

{[}44{]} P.~Caucal, E.~Iancu, A.~H.~Mueller, and G.~Soyez, ``Vacuum-like
jet fragmentation in a dense QCD medium,'' Phys. Rev. Lett.
\textbf{120}, 232001 (2018), arXiv:1801.09703.

{[}45{]} Y.~Mehtar-Tani, D.~Pablos, and K.~Tywoniuk, ``Jet suppression
and azimuthal anisotropy from RHIC to LHC,'' Phys. Rev. D \textbf{110},
014009 (2024), arXiv:2402.07869.

{[}46{]} R.~Ehlers \emph{et al.}, ``Bayesian inference analysis of jet
quenching using inclusive jet and hadron suppression measurements,''
Phys. Rev. C \textbf{111}, 054913 (2025), arXiv:2408.08247.

{[}47{]} J.~Casalderrey-Solana, D.~C.~Gulhan, J.~G.~Milhano, D.~Pablos,
and K.~Rajagopal, ``A hybrid strong/weak coupling approach to jet
quenching,'' JHEP \textbf{10}, 019 (2014) {[}Erratum: JHEP 09, 175
(2015){]}, arXiv:1405.3864.

\end{document}